%% file: main.tex
\pdfoutput=1   

\PassOptionsToPackage{table}{xcolor}
\documentclass{article}
\usepackage{preprint,times}

\input{math_commands.tex}

\usepackage{hyperref}
\usepackage{url}
\usepackage[labelsep=period]{caption}
\setcitestyle{citesep={,},yysep={,}}

\usepackage{graphicx}
\usepackage{booktabs}
\usepackage{xcolor}
\usepackage{pifont}
\usepackage{amssymb}
\usepackage{multirow}
\usepackage{enumitem}
\usepackage{flafter}
\usepackage{placeins}

\definecolor{active}{RGB}{200,230,255}
\definecolor{inactive}{RGB}{240,240,240}
\definecolor{tabbanda}{gray}{1.00}
\definecolor{tabbandb}{gray}{0.93}
\definecolor{realtime}{RGB}{220,240,220}

\title{EL{\normalfont i}Cv2: Learned LiDAR Geometry Compression under Real-Time Embedded Constraints}

\author{
Soowoong Kim \quad Junsik Kim \quad Gun Bang \quad Jinsoo Choi \\
Electronics and Telecommunications Research Institute (ETRI) \\
Daejeon, Republic of Korea \\
\texttt{\{soowoong.kim, junsikkim, gbang, jschoi\}@etri.re.kr}
}

\hypersetup{
  pdftitle={ELiCv2: Learned LiDAR Geometry Compression under Real-Time Embedded Constraints},
  pdfauthor={Soowoong Kim, Junsik Kim, Gun Bang, Jinsoo Choi},
  pdfsubject={Learned point cloud compression},
  pdfkeywords={LiDAR compression, point cloud compression, learned compression,
               entropy model, sparse convolution, embedded inference}
}

\ppfinalcopy   

\begin{document}

\maketitle
\lhead{}\rhead{}\chead{}   

\input{0_abstract_arxiv}
\input{1_intro}
\input{2_related}
\input{3_method}
\input{4_exp}
\input{5_outro}

\section*{AI Assistance Disclosure}
We used ChatGPT to assist with writing and polishing the manuscript,
and Claude to assist with creating some of the figures.

\bibliography{refs}
\bibliographystyle{preprint}

\clearpage
\appendix

\input{app_split}
\FloatBarrier
\input{app_entropy}
\FloatBarrier
\input{app_table_size}
\FloatBarrier
\input{app_kmap}
\FloatBarrier
\input{app_large_kmap}
\FloatBarrier
\input{app_inter_align}
\FloatBarrier
\input{app_why}
\FloatBarrier
\input{app_energy}
\FloatBarrier
\input{app_memory}
\FloatBarrier

\end{document}

%% file: math_commands.tex
\usepackage{amsmath,amsfonts,bm}

\def\eqref#1{equation~\ref{#1}}

\def\1{\bm{1}}

\DeclareMathAlphabet{\mathsfit}{\encodingdefault}{\sfdefault}{m}{sl}
\SetMathAlphabet{\mathsfit}{bold}{\encodingdefault}{\sfdefault}{bx}{n}



%% file: 0_abstract_arxiv.tex
\begin{abstract}
Improving learned LiDAR compression increasingly relies on heavier inference,
making it harder to meet the constraints of real-time embedded systems. We present
ELiCv2, which removes two sources of redundant computation in progressive
coding. Its discrete entropy interface selects a precomputed integer CDF
instead of materializing one from a per-voxel distribution, while its
occupancy-driven kernel-map builder derives finer-level sparse-convolution
kernel-maps directly from the decoded hierarchy, eliminating coordinate-based
neighbor search. ELiCv2 reinvests the recovered computation budget in
lightweight ego-motion-aware temporal conditioning that gathers local occupancy
context from the aligned previous reconstruction. On four LiDAR datasets and an NVIDIA
Jetson Thor, ELiCv2
achieves a mean BD-rate reduction of $30.97\%$ over G-PCC with channel width
$C{=}32$ while sustaining the 20Hz nuScenes capture rate. Increasing the
channel width to $C{=}64$ improves the mean reduction to $35.22\%$.
\end{abstract}

%% file: 1_intro.tex
\section{Introduction}
\label{sec:intro}

LiDAR geometry compression is increasingly important for applications such as
autonomous driving, robotics, V2X, digital twins, and large-scale 3D
mapping~\citep{mounier2024lidar,Lin_2025_CVPR,wang2025v2pnet,wang2026pvf,guang2026mdslca}.
These applications produce high-rate point clouds that must be transmitted
under strict bandwidth and computational constraints. In practical edge--server
deployments, LiDAR data are captured on resource-constrained platforms such as
robots, drones, and autonomous vehicles, while downstream perception is often
executed on more capable servers~\citep{zhang2021emp,luo2023edgecooper}. This asymmetry calls for lightweight encoders
that can operate under limited compute budgets while keeping pace with
high-rate LiDAR sensors.

Recent learned codecs substantially improve compression efficiency.
Hybrid voxel--octree methods such as RENO~\citep{RENO} and
ELiC~\citep{ELiC} balance compression performance and runtime,
achieving real-time throughput on desktop GPUs.
Other approaches, including TopNet~\citep{Wang_2025_CVPR},
AnyPcc~\citep{wang2026anypcc}, and Unicorn~\citep{unicorn},
exploit attention-based or multistage context modeling to improve
compression efficiency.
Despite these advances, achieving strong compression performance
while sustaining real-time encoding under embedded compute
constraints remains challenging.

This challenge motivates us to examine how learned codecs spend
their limited computation budget.
Beyond context prediction, progressive occupancy coding also incurs
costs when converting network outputs into coding distributions and
constructing sparse-convolution kernel-maps.
A conventional neural entropy interface converts each categorical
output into an integer CDF for entropy coding.
Sparse-convolution frameworks such as SpConv, TorchSparse++, and
MinkowskiEngine construct kernel-maps through coordinate-based
lookup~\citep{spconv2022,torchsparse,minkowskiengine}, even when
decoded occupancy and the parent kernel-map already determine
the child kernel-map.
These operations suggest an opportunity to recover computation
for prediction by eliminating repeated distribution conversion
and exploiting the hierarchy exposed by progressive decoding.

ELiCv2 eliminates both forms of avoidable computation.
The network directly selects a coder-ready distribution from
precomputed integer CDFs, while decoded occupancy and the parent
kernel-map determine the child kernel-map without coordinate
search at finer levels.
Together, these mechanisms reduce the time spent preparing
entropy-coder inputs and constructing kernel-maps, leaving more
of the runtime budget available for context prediction.
ELiCv2 uses this budget to exploit redundancy between consecutive
LiDAR frames through a lightweight local gate conditioned on the
ego-motion-aligned previous decoded frame.
This adds temporal context without motion search or a separate
temporal backbone.

\begin{figure}[h]
\centering
\includegraphics[width=\linewidth,trim={0 6bp 0 6bp},clip]{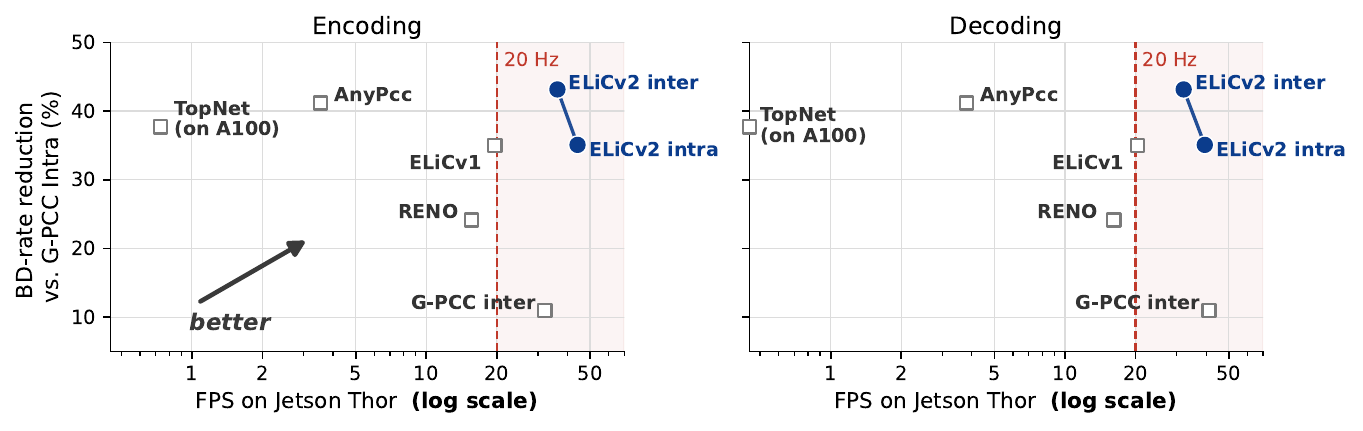}
\caption{Compression efficiency versus throughput on an NVIDIA Jetson Thor.
ELiCv2 reallocates computation from avoidable coding overhead to temporal context while sustaining the 20Hz nuScenes sensor rate.
Throughput is measured at bit-depth 14, and BD-rate over bit-depths ${16,15,14,13,12}$.}
\label{fig:teaser}
\end{figure}

As shown in Figure~\ref{fig:teaser}, ELiCv2 achieves the fastest learned
encoding among the evaluated methods on nuScenes using an NVIDIA
Jetson Thor. Its $C{=}32$ inter model sustains the 20Hz sensor rate
and achieves a $43.15\%$ BD-rate reduction relative to G-PCC intra
on this dataset.

Our main contributions are listed below.
\begin{itemize}[leftmargin=*, topsep=2pt, itemsep=1pt, parsep=0pt]
    \item We identify two sources of avoidable computation in the entropy and
    sparse-processing interfaces of progressive coding.

    \item We replace per-voxel CDF materialization and coordinate search with
    direct selection of precomputed integer CDFs and occupancy-driven
    kernel-map propagation.

    \item We reinvest the recovered budget in lightweight ego-motion-aware
    temporal coding that gathers local context from the aligned previous
    decoded frame without motion search or complex temporal processing.
\end{itemize}

%% file: 2_related.tex
\section{Related Work}
\label{sec:related}

\paragraph{Learned LiDAR geometry compression.}
Learned point cloud geometry codecs span range-image, octree, and sparse-voxel
representations. Range-image methods~\citep{zhou2022riddle,liu2023bird,zhao2022real}
exploit efficient 2D processing but can lose geometric information through
projection. Octree-based methods~\citep{OctAttention,ECMOPCC,VoxelContextNet,Wang_2025_CVPR}
model hierarchical occupancy with increasingly powerful context models, achieving
strong compression at the cost of sequential or autoregressive inference.
Sparse-convolutional methods~\citep{SparsePCGC,UniPCGC} operate directly on
voxelized geometry and provide greater parallelism, although sparse processing
can introduce substantial computational overhead.

Hybrid voxel--octree codecs such as RENO~\citep{RENO} and ELiC~\citep{ELiC}
combine sparse-convolutional context modeling with progressive occupancy coding,
providing a favorable balance between compression efficiency and runtime.
ELiCv2 builds on this direction while explicitly targeting embedded inference.
It replaces generic entropy-coding and sparse-convolution interfaces with
coder-ready table selection and hierarchy-aware kernel-map propagation
rather than increasing entropy-model complexity.

\paragraph{Inter-frame point cloud compression.}
Learned inter-frame codecs exploit temporal redundancy through motion estimation
or cross-frame feature alignment. D-DPCC~\citep{ddpcc} predicts
feature-space motion and interpolates reference features, while
Akhtar et al.~\citep{akhtar2024} learn cross-frame correspondence using generalized
sparse convolution. Both primarily target dense dynamic point clouds.
Unicorn~\citep{unicorn} supports inter-frame geometry coding
for diverse point clouds, including LiDAR, using a multiscale
reference pyramid and sparse target convolutions.
However, its wide $9\!\times\!9\!\times\!9$ cross-frame operators incur substantial inference cost.

RIDDLE~\citep{zhou2022riddle} incorporates sensor poses into
temporal range prediction by querying neighboring points from
the previous scan and reprojecting them into the current laser
shot's coordinate system.

ELiCv2 uses ego-motion-aligned references within progressive
occupancy coding. Since the current-level voxel coordinates
are already known, temporal context is gathered from fixed
$3\!\times\!3\!\times\!3$ neighborhoods in the aligned reference
occupancy pyramid and injected through a lightweight local gate.
This avoids the estimated query positions and point-neighbor
search used by RIDDLE, while preserving parallel processing
across voxels within each coding stage.

\paragraph{Kernel-map construction for sparse convolution.}
Sparse convolution requires a kernel-map specifying neighboring active voxels
for each kernel offset. General-purpose engines such as
MinkowskiEngine~\citep{minkowskiengine}, SpConv~\citep{spconv2022}, and
TorchSparse++~\citep{torchsparse} construct these maps from sparse coordinates,
typically through hashing or coordinate search. These engines support arbitrary
sparse tensors and therefore reconstruct kernel-maps as active coordinates
change. Progressive occupancy coding provides additional structure because finer-level
voxels are deterministically generated from decoded parent occupancies.
ELiCv2 exploits this hierarchy to propagate kernel-maps across octree levels,
avoiding repeated coordinate-based connectivity search.

%% file: 3_method.tex
\section{Method}
\label{sec:method}

\subsection{Progressive Structure and Design Principle}
\label{sec:method-prelim}

ELiCv2 builds upon the progressive occupancy-coding framework
of ELiC~\citep{ELiC}, retaining its sparse-convolutional backbone
and cross-bit-depth feature propagation.
A point cloud quantized to bit-depth $B$ is represented as an octree and encoded
coarse-to-fine over levels $\ell=2,\ldots,B-1$.
At each level, an occupied voxel $v\in\mathcal{V}_\ell$ is represented by an 8-bit
occupancy symbol $o_v\in\{1,\ldots,255\}$ indicating its occupied children.
Once level $\ell$ is decoded, the coordinates at level $\ell+1$ are reconstructed
deterministically, providing identical multiscale contexts to the encoder and decoder.

Following ELiC, each occupancy symbol is factorized into two sequential 4-bit
symbols, reducing entropy prediction to two 16-way classification problems.
The two stages are sequential, but all voxels within each stage are processed
in parallel rather than through a voxel-by-voxel autoregressive chain.
Unlike the low/high nibble split of ELiC, we partition the eight
octants according to the parity of $x+y+z$.
For octant index $i=4z+2y+x$, with $(x,y,z)\in\{0,1\}^3$, we use
\begin{equation}
    s_0=\{1,2,4,7\},
    \qquad
    s_1=\{0,3,5,6\}.
    \label{eq:octant-parity-split}
\end{equation}
This octant parity partition makes each second-stage octant face-adjacent to
three first-stage octants, providing local conditioning without additional
coding stages. Appendix~\ref{app:split} shows that it reduces rate at all 20
evaluated dataset and bit-depth combinations relative to the planar split,
with a mean reduction of $0.125$ bpp.

Progressive decoding exposes discrete structure that generic neural interfaces
do not preserve. Once an occupancy symbol is decoded, the child support and
parent--child relations are known deterministically. ELiCv2 uses this hierarchy
to derive finer-level kernel-maps without coordinate search. The same principle
applies at the entropy interface, where the network directly selects a
coder-ready integer CDF instead of materializing one from each categorical
output. By removing these two sources of interface overhead without reducing
the ELiC backbone, ELiCv2 recovers computation for lightweight temporal
context.

\begin{figure}[t]
    \centering
    \includegraphics[width=\linewidth,trim={0 7bp 0 7bp},clip]{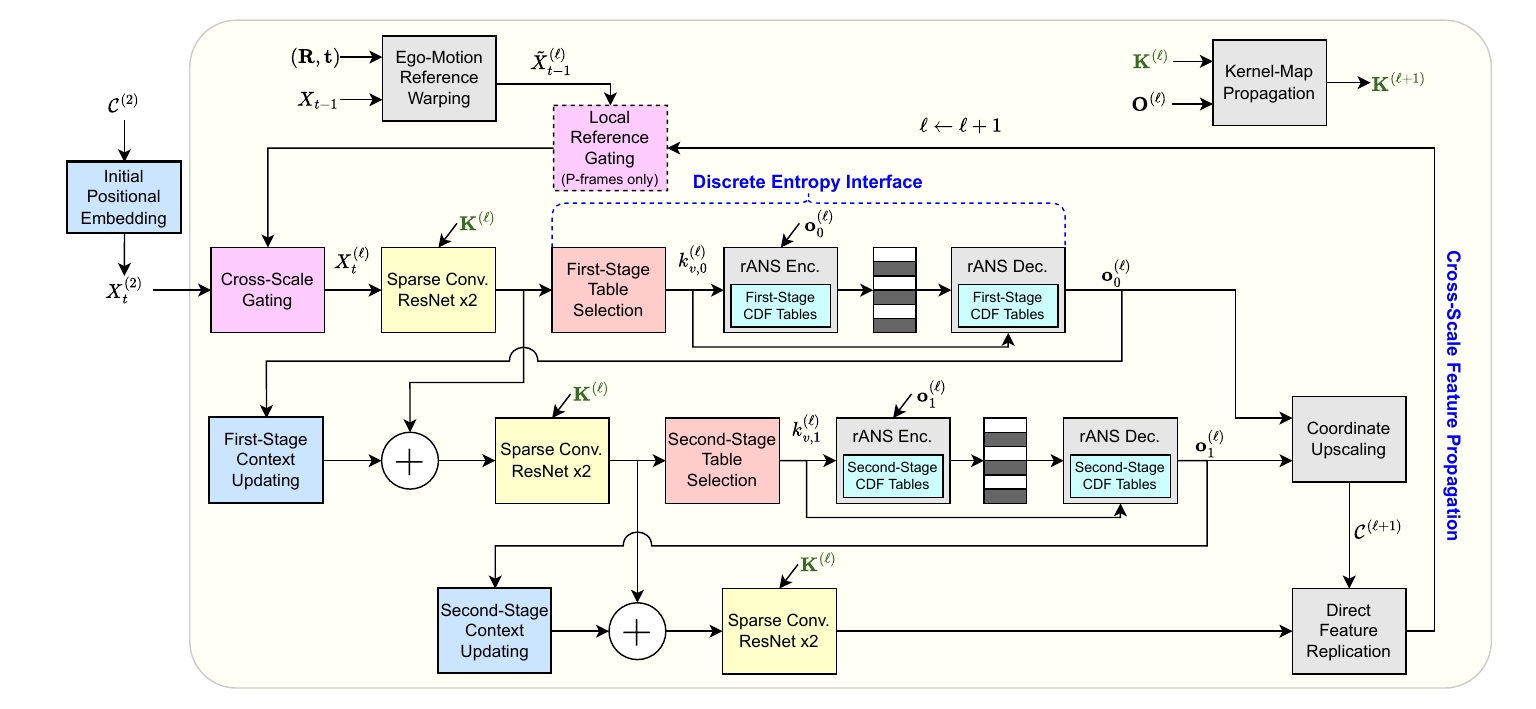}
    \caption{
    Overview of ELiCv2. At each octree level, the backbone selects indexed CDF
    tables for two-stage occupancy coding. The decoded occupancy and current
    kernel-map determine the next-level kernel-map. P-frames additionally use
    local gating over an ego-motion-aligned reference.
    }
    \label{fig:elicv2-overview}
\end{figure}

Figure~\ref{fig:elicv2-overview} summarizes how the discrete entropy interface,
occupancy-driven kernel-map propagation, and ego-motion-aligned reference
gating are integrated into the progressive backbone.

\subsection{Discrete Entropy Interface}
\label{sec:method-table}

Conventional neural entropy models output a categorical probability distribution
over occupancy symbols for each voxel. Before encoding or decoding a symbol,
these probabilities must be quantized into integer frequencies and accumulated
into a finite-precision cumulative distribution function (CDF) required by the
entropy coder. This
probability-to-coder materialization incurs substantial preparation and
data-transfer overhead even though rANS~\citep{duda2015use} ultimately consumes a discrete coding
rule. ELiCv2 redesigns this interface. It learns a finite codebook of
distributions and predicts which coder-ready CDF table should be used.

We retain the Bag-of-Encoders (BoE) scheme of ELiCv1~\citep{ELiC},
using a total of $K_{\mathrm{BoE}}{=}4$ coding networks. BoE selects a coding
network at each bit depth according to occupancy statistics. The selected network
then predicts a CDF-table index for each node.

Within each coding network, the codebook is parameterized using row and column
sets of $T$ learnable logits,
$\mathbf{L}^{\mathrm{row}},\mathbf{L}^{\mathrm{col}}\in\mathbb{R}^{T\times16}$, low-rank interaction
factors
$\boldsymbol{\alpha}^{\mathrm{row}},\boldsymbol{\alpha}^{\mathrm{col}}\in\mathbb{R}^{T\times r}$,
and a shared basis $\mathbf{B}\in\mathbb{R}^{r\times16}$.
Each pair $(i,j)$ defines
\begin{equation}
    \mathbf{p}_{ij}
    =
    \operatorname{softmax}\!\left(
        \mathbf{L}^{\mathrm{row}}_{i}
        +
        \mathbf{L}^{\mathrm{col}}_{j}
        +
        (\boldsymbol{\alpha}^{\mathrm{row}}_{i}
        \odot
        \boldsymbol{\alpha}^{\mathrm{col}}_{j})\mathbf{B}
    \right).
    \label{eq:pair-pmf}
\end{equation}
The low-rank interaction provides $T^2$ expressive distributions without
independently parameterizing all table pairs.

For each voxel $v$, a two-layer MLP selection head outputs
$2T$ logits, grouped into row and column vectors
$\mathbf{z}^{\mathrm{row}}_v,\mathbf{z}^{\mathrm{col}}_v\in\mathbb{R}^{T}$.
The table indices are selected as
\begin{equation}
    i_v=\operatorname*{argmax}_{i}\mathbf{z}^{\mathrm{row}}_{v,i},
    \qquad
    j_v=\operatorname*{argmax}_{j}\mathbf{z}^{\mathrm{col}}_{v,j},
    \qquad
    k_v=T i_v+j_v ,
    \label{eq:table-select}
\end{equation}
where $i_v,j_v\in\{0,\ldots,T-1\}$ are zero-based row and column indices,
and $k_v\in\{0,\ldots,T^2-1\}$ is the flattened index identifying the
selected distribution $\mathbf{p}_{i_vj_v}$ for voxel $v$.
After training, all $T^2$ distributions are quantized into fixed integer CDFs
shared by the encoder and decoder. The index $k_v$ identifies the CDF used
by indexed rANS to encode or decode the occupancy symbol of voxel $v$
for the current coding network and stage.
With $T=32$, this gives $1024$ CDF tables.
Appendix~\ref{app:table-size} examines the rate--storage trade-off underlying
this choice.
At inference, the network therefore selects a discrete coding distribution
directly. Entropy preparation becomes top-1 selection and an integer table
lookup followed by indexed rANS coding, eliminating per-voxel CDF
materialization.
We use a Gumbel--Softmax relaxation~\citep{jang2016categorical}
during training and hard top-1 selection at inference.

\subsection{Occupancy-Driven Kernel-Map Propagation}
\label{sec:method-kmap}

General sparse engines assume that each active support is arbitrary. They
therefore recover $3\!\times\!3\!\times\!3$ connectivity through coordinate hashing or
neighbor search whenever the support changes. This generic interface discards
the hierarchy exposed by progressive occupancy coding. Every finer-level voxel
is generated from a known parent according to decoded occupancy, so child
connectivity is already determined by parent connectivity and occupancy masks.
ELiCv2 reformulates kernel-map construction from coordinate search into
hierarchical kernel-map propagation.

Let $\mathbf{K}^{p}\in\mathbb{Z}^{N_p\times27}$ be the parent kernel-map and
$m_i\in\{0,\ldots,255\}$ the 8-bit integer occupancy mask of parent $i$.
We use $m_i[q]\in\{0,1\}$ to denote its bit at octant $q$.
An active child in octant $q$ receives the index
\begin{equation}
    \operatorname{idx}(i,q)
    =
    \operatorname{base}(i)
    +
    \operatorname{popcount}
    \left(m_i\wedge(2^q-1)\right),
    \label{eq:child-idx}
\end{equation}
where $\operatorname{base}(i)$ is an exclusive prefix sum over the numbers of
active children.

For each child octant $q$ and kernel offset $k$, the corresponding neighboring
parent and target child octant depend only on $(q,k)$. We precompute these
relations as $T_{\mathrm{col}}(q,k)$ and $T_{\mathrm{bit}}(q,k)$.
The child kernel-map is then obtained directly from the parent map and occupancy.
\begin{equation}
    \mathbf{K}^{c}[u,k]
    =
    \begin{cases}
        \operatorname{idx}(j,q'),
        & j=\mathbf{K}^{p}[i,T_{\mathrm{col}}(q,k)]\geq0
          \ \text{and}\ m_j[q']=1,\\
        -1, & \text{otherwise},
    \end{cases}
    \label{eq:kmap-propagation}
\end{equation}
where $u=\operatorname{idx}(i,q)$ and
$q'=T_{\mathrm{bit}}(q,k)$.

Thus, the progressive coarse-to-fine structure turns kernel-map construction
from coordinate search into deterministic kernel-map propagation using only
prefix sums, table lookups, and bit operations. The key improvement is not
faster search. Coordinate search is unnecessary at progressively generated
finer levels and is eliminated entirely at those levels.
The same principle is not specific to the $3\!\times\!3\!\times\!3$ stencil used in
ELiCv2 and extends to larger kernel neighborhoods, as demonstrated in
Appendix~\ref{app:kmap-runtime}. The full derivation of the propagation rule
is provided in Appendix~\ref{app:rulebook}.

\subsection{Ego-Motion-Aware Inter Coding}
\label{sec:method-inter}

After eliminating redundant entropy and sparse-processing work, we reinvest
the recovered computation budget in predictive context. For a P-frame, the
previous reconstructed frame is aligned to the current frame using sensor
ego-motion and used as additional context. Since geometry coding is lossless,
the encoder already holds the same previous reconstruction available to the
decoder, requiring neither closed-loop decoding nor an additional temporal
backbone.

\paragraph{Reference construction.}
Let $X_t$ and $X_{t-1}$ denote the current and previous voxelized frames.
We align $X_{t-1}$ to the current coordinate system as
\begin{equation}
    \tilde{X}_{t-1}
    =
    \mathcal{W}_{t-1\rightarrow t}(X_{t-1}),
    \qquad
    \mathcal{W}(\mathbf{x})
    =
    \left\lfloor
    \frac{\mathbf{R}\mathbf{x}+\mathbf{t}+2^{29}}{2^{30}}
    \right\rfloor ,
    \label{eq:warp}
\end{equation}
where $(\mathbf{R},\mathbf{t})$ is the relative ego transform quantized to
30-bit fixed-point parameters
and transmitted in the bitstream. Both encoder and decoder therefore construct
an identical reference using integer arithmetic.
The reported inter-frame bitrates include the uncompressed 96-byte P-frame
header containing the nine entries of $\mathbf{R}$ and three entries of
$\mathbf{t}$ as 64-bit integers.
The aligned frame is
downsampled to form the multiscale reference pyramid
$\tilde{X}^{(\ell)}=(\tilde{\mathcal{C}}^{(\ell)},\tilde{O}^{(\ell)})$.

\paragraph{Local reference gating.}
Exact voxel correspondence is unreliable because consecutive LiDAR sweeps
sample slightly different spatial locations. For each current voxel, we
therefore gather its $3\!\times\!3\!\times\!3$ reference neighborhood.
\begin{equation}
    r_{v,i}
    =
    \begin{cases}
        \tilde{O}^{(\ell)}[\mathbf{x}_v+\boldsymbol{\delta}_i],
        & \mathbf{x}_v+\boldsymbol{\delta}_i\in\tilde{\mathcal{C}}^{(\ell)},\\
        \textsc{absent},
        & \text{otherwise},
    \end{cases}
    \qquad i=1,\ldots,27.
    \label{eq:refgather}
\end{equation}
The gathered symbols are embedded and combined into a position-aware context,
\begin{equation}
    \mathbf{h}^{\mathrm{ref}}_v
    =
    \mathbf{W}
    \left[
        E(r_{v,1})\Vert\cdots\Vert E(r_{v,27})
    \right]
    + \mathbf{b},
    \label{eq:refctx}
\end{equation}
where $\Vert$ denotes feature concatenation and $E$ is a learned
embedding table shared across all $27$ offsets, with a dedicated entry
for \textsc{absent}. Each block $\mathbf{W}_i$ of the projection
$\mathbf{W}$ corresponds to one offset, preserving the relative spatial
position of the embedded reference symbol. The embedding table and bias
$\mathbf{b}$ are initialized to zero, so the temporal path initially
contributes no reference context.
The resulting context is injected into the propagated feature through
\begin{equation}
    \mathbf{f}_v
    \leftarrow
    \mathbf{f}_v
    +
    \sigma\!\left(g([\mathbf{f}_v\Vert \mathbf{h}^{\mathrm{ref}}_v])\right)
    \odot \mathbf{h}^{\mathrm{ref}}_v .
    \label{eq:refgate}
\end{equation}
where $g$ is a two-layer MLP and $\sigma$ denotes the sigmoid function.
This position-aware gate is the only structural addition to the intra
backbone. Temporal conditioning is restricted to
$\ell\leq \ell_{\mathrm{ref}}=13$, beyond which voxel-level correspondence rapidly
deteriorates. Appendix~\ref{app:inter-align} provides the alignment
visualization and resolution-dependent analysis.

Overall, the inter model is both the payoff of structure-preserving inference
and a lightweight temporal mechanism. It exploits pose-aligned redundancy
using only local occupancy gathering and feature gating without motion search
or complex temporal processing. Further design rationale and comparisons with
cross-frame operators are provided in Appendix~\ref{app:why-alignment}.

%% file: 4_exp.tex
\section{Experiments}
\label{sec:experiments}

\subsection{Experimental Setup}
\label{sec:experimental-setup}

\paragraph{Datasets.}
We evaluate ELiCv2 on four LiDAR sequence datasets.
SemanticKITTI~\citep{semantickitti}, Ford~\citep{ford},
nuScenes~\citep{caesar2020nuscenes}, and V2X-Seq~\citep{v2x-seq}.
These datasets cover diverse acquisition settings, from moving vehicle-mounted
LiDAR to stationary infrastructure LiDAR, allowing us to evaluate inter-frame
coding under different ego-motion characteristics.
Table~\ref{tab:datasets} summarizes the dataset statistics.

\begin{table}[t]
\centering
\small
\caption{Dataset statistics. Points denotes the mean number of points per test
frame, and ego step is the median translation between consecutive sensor poses.}
\label{tab:datasets}
\begin{tabular}{lrrrrl}
\toprule
Dataset & Train & Test & Points & Ego step & Setting \\
\midrule
\rowcolor{tabbanda}
SemanticKITTI & 23{,}201  & 20{,}351 & 120{,}557 & $0.932$\,m & Vehicle, 10Hz \\
\rowcolor{tabbandb}
Ford          & 1{,}500   & 3{,}000  & 83{,}927  & $0.634$\,m & Vehicle \\
\rowcolor{tabbanda}
nuScenes      & 297{,}737 & 52{,}423 & 34{,}722  & $0.314$\,m & Vehicle, 20Hz \\
\rowcolor{tabbandb}
V2X-Seq       & 11{,}275  & 4{,}862  & 56{,}273  & $0$         & Infrastructure \\
\bottomrule
\end{tabular}
\end{table}

For SemanticKITTI, we use sequences $00$--$10$ for training and $11$--$21$
for testing, with sensor poses obtained from the provided odometry and
calibration. For Ford, we train on Ford\_01 and evaluate on Ford\_02 and
Ford\_03. Since ground-truth sensor poses are unavailable, inter-frame
transforms are estimated using KISS-ICP~\citep{vizzo2023kiss}. For nuScenes, we use the full 20Hz
LiDAR sweep stream rather than the 2Hz annotated keyframes, with LiDAR
poses obtained by composing the provided ego poses and sensor extrinsics.
V2X-Seq consists of stationary roadside LiDAR, for which the relative sensor
transform is the identity.

\paragraph{Training.}
We train a separate model for each dataset at the finest operating point
($\mathrm{posQ}=4$, corresponding to a 16-bit octree) and use the same model for all
coarser rate points. Since ELiCv2 performs lossless geometry coding on the
quantized voxel grid, no distortion term is required and the training objective
directly minimizes the expected code length. Specifically, for a voxel $v$ at
octree level $\ell$ and coding stage $s\in\{0,1\}$, we minimize
\begin{equation}
    \mathcal{L}
    =
    \frac{1}{N}
    \sum_{\ell}
    \sum_{v\in\mathcal{V}_{\ell}}
    \sum_{s\in\{0,1\}}
    -\log_{2} \mathbf{p}_{v,s}\!\left[o_{v,s}\right],
    \label{eq:training-loss}
\end{equation}
where $o_{v,s}\in\{0,\ldots,15\}$ is the target 4-bit occupancy symbol,
$\mathbf{p}_{v,s}$ is the corresponding 16-way distribution produced by the entropy
model, and $N$ is the number of input points.

The intra model is trained for $300$k steps using Adam~\citep{kingma2014adam} with a batch size of
$4$ and a cosine learning-rate schedule from $5\times10^{-4}$ to
$5\times10^{-6}$. Since the hard CDF-table selection in
Section~\ref{sec:method-table} is non-differentiable, we use a
Gumbel--Softmax relaxation~\citep{jang2016categorical} during training and anneal its temperature from
$\tau=2.0$ to $\tau=0.1$.

The inter model is initialized from the corresponding trained intra model and
jointly fine-tuned for $200$k steps. We use Adam with a cosine learning-rate
schedule from $1\times10^{-4}$ to $5\times10^{-6}$ and fix the
Gumbel--Softmax temperature at $\tau=0.1$. Each training sample consists of
consecutive LiDAR frames, with four P-frames conditioned on their respective
preceding frames. 

\paragraph{Evaluation metrics.}
We evaluate compression performance using bits per input point (bpp) and
D1-PSNR. For all operating points, bpp is normalized by the number of occupied
voxels at the finest quantization level ($\mathrm{posQ}=1$), providing a fixed
denominator across rate points. Since all evaluated codecs are lossless on the
quantized voxel grid, distortion at a given bit-depth is determined solely by
quantization and is identical across methods.

We summarize rate--distortion performance using BD-rate~\citep{bdrate} with
G-PCC intra as the anchor. Unless otherwise stated, compression results are
evaluated over the complete test split of each dataset.

For inter-frame coding, we use a GOP size of $30$ with an IPPPP structure.
The first frame is coded by the intra model, and each of the following
$29$ P-frames references its immediate predecessor. GOPs never cross sequence
boundaries.

\paragraph{Implementation and hardware.}
ELiCv2 is implemented in PyTorch~\citep{pytorch} with sparse-convolution operations based on
TorchSparse++~\citep{torchsparse}, and uses 16-bit rANS for entropy coding.
The conventional baselines use the G-PCC standard~\citep{GPCC} and its TM13v34 test model~\citep{GPCCv34}.
All models are evaluated in FP32 and trained on NVIDIA A100 GPUs. Runtime
measurements are conducted on an NVIDIA Jetson Thor~\citep{jetsonthor}, our
target embedded platform, under the 120~W nvpmodel power profile, with the CPU
and GPU clocks fixed at 2,601~MHz and 1,386~MHz, respectively. Encoding and decoding latency excludes file I/O. All methods use
the same Jetson Thor hardware and power settings, except TopNet,
which is measured on an A100 because it does not fit on the Jetson.

\subsection{Compression Performance}
\label{sec:compression-performance}

Table~\ref{tab:bdrate-all} compares the compression performance of ELiCv2
with TopNet~\citep{Wang_2025_CVPR}, RENO~\citep{RENO},
AnyPcc~\citep{wang2026anypcc}, and ELiCv1~\citep{ELiC}, together with the
G-PCC baselines. The $C{=}32$ intra model
achieves a mean BD-rate reduction of $26.66\%$ over G-PCC, which improves to
$30.97\%$ with inter coding. Increasing the model capacity to $C{=}64$ further
improves the mean reduction from $31.47\%$ for intra coding to $35.22\%$ for
inter coding.

\begin{table}[h]
\centering
\small
\setlength{\tabcolsep}{4pt}
\renewcommand{\arraystretch}{1.15}
\caption{BD-rate (\%) relative to G-PCC intra over bit-depths
$\{16,15,14,13,12\}$. Negative values indicate better compression.
$^\dagger$ TopNet is evaluated on a 600-frame sample. Best results are in
\textbf{bold}.}
\label{tab:bdrate-all}
\begin{tabular}{lrrrrr}
\toprule
Method & nuScenes & Sem.KITTI & Ford & V2X-Seq & Mean \\
\midrule
\rowcolor{tabbanda}
G-PCC (intra)
& 0.00 & 0.00 & 0.00 & 0.00 & 0.00 \\

\rowcolor{tabbandb}
G-PCC (inter)
& -10.98 & -1.64 & -4.80 & -4.25 & -5.42 \\

\rowcolor{tabbanda}
TopNet$^\dagger$
& -37.71 & -31.66 & \textbf{-30.00} & -19.59 & -29.74 \\

\rowcolor{tabbandb}
RENO
& -24.16 & -17.11 & -14.53 & -12.28 & -17.02 \\

\rowcolor{tabbanda}
AnyPcc
& -41.19 & -32.57 & -28.32 & -29.40 & -32.87 \\

\rowcolor{tabbandb}
ELiCv1
& -34.99 & -26.06 & -23.72 & -25.40 & -27.54 \\

\midrule

\rowcolor{tabbanda}
Ours ($C{=}32$, intra)
& -35.08 & -24.62 & -23.57 & -23.39 & -26.66 \\

\rowcolor{tabbandb}
Ours ($C{=}32$, inter)
& -43.15 & -28.27 & -27.35 & -25.09 & -30.97 \\

\rowcolor{tabbanda}
Ours ($C{=}64$, intra)
& -40.47 & -29.09 & -27.03 & -29.27 & -31.47 \\

\rowcolor{tabbandb}
Ours ($C{=}64$, inter)
& \textbf{-48.24} & \textbf{-32.61} & -29.73 &
\textbf{-30.32} & \textbf{-35.22} \\

\bottomrule
\end{tabular}
\end{table}

The $C{=}32$ configuration provides a strong operating point for embedded
deployment. Its inter model achieves a mean BD-rate reduction of $30.97\%$,
outperforming ELiCv1 ($27.54\%$), RENO ($17.02\%$), and the sampled TopNet
result ($29.74\%$). The improvement is particularly large on nuScenes, where
ELiCv2 $C{=}32$ inter reduces BD-rate by $43.15\%$. As shown in
Section~\ref{sec:experiments-runtime}, these gains are obtained while retaining
the low latency targeted by ELiCv2.

The larger $C{=}64$ model provides a compression-oriented operating point.
With inter coding, it achieves the best results on nuScenes ($-48.24\%$),
SemanticKITTI ($-32.61\%$), and V2X-Seq ($-30.32\%$), while remaining
competitive with TopNet on Ford. Its mean BD-rate reaches $-35.22\%$,
improving upon AnyPcc ($-32.87\%$), the strongest competing method on average.

Inter coding consistently improves both model capacities across all four
datasets. Relative to their intra models, the $C{=}32$ and
$C{=}64$ inter models provide additional mean BD-rate reductions of $4.31$
and $3.75$ percentage points, respectively. This consistent improvement shows
the payoff of computation reallocation. After removing entropy preparation and
kernel-map construction overhead, ELiCv2 adds useful temporal context without
motion search or complex temporal processing.

\subsection{Runtime Performance}
\label{sec:experiments-runtime}

\begin{table}[!ht]
\centering
\scriptsize
\setlength{\tabcolsep}{2pt}
\renewcommand{\arraystretch}{1.15}
\caption{
Encoding and decoding latency (ms/frame) on NVIDIA Jetson Thor with I/O excluded.
Green cells indicate real-time operation at the corresponding sensor rate
(20Hz for nuScenes and 10Hz for the other datasets).
$^\dagger$ TopNet is measured on an A100 because it does not fit on the Jetson.
}
\label{tab:latency-all}

\resizebox{\linewidth}{!}{%
\begin{tabular}{llrrrrrrrrrrrrrrrrrrrr}
\toprule
& & \multicolumn{5}{c}{nuScenes}
& \multicolumn{5}{c}{SemanticKITTI}
& \multicolumn{5}{c}{Ford}
& \multicolumn{5}{c}{V2X-Seq} \\
\cmidrule(lr){3-7}
\cmidrule(lr){8-12}
\cmidrule(lr){13-17}
\cmidrule(lr){18-22}
Method &
& 16 & 15 & 14 & 13 & 12& 16 & 15 & 14 & 13 & 12& 16 & 15 & 14 & 13 & 12& 16 & 15 & 14 & 13 & 12 \\
\midrule

\rowcolor{tabbanda} & enc & \cellcolor{realtime}38.7 & \cellcolor{realtime}31.1 & \cellcolor{realtime}24.7 & \cellcolor{realtime}19.3 & \cellcolor{realtime}13.7 & 140.0 & 112.5 & \cellcolor{realtime}83.4 & \cellcolor{realtime}54.8 & \cellcolor{realtime}31.2 & 107.6 & \cellcolor{realtime}91.2 & \cellcolor{realtime}72.9 & \cellcolor{realtime}55.6 & \cellcolor{realtime}38.0 & \cellcolor{realtime}90.8 & \cellcolor{realtime}78.4 & \cellcolor{realtime}65.9 & \cellcolor{realtime}52.8 & \cellcolor{realtime}40.1 \\
\rowcolor{tabbanda}\multirow{-2}{*}{G-PCC (intra)} & dec & \cellcolor{realtime}28.3 & \cellcolor{realtime}22.7 & \cellcolor{realtime}17.7 & \cellcolor{realtime}13.1 & \cellcolor{realtime}8.8 & 103.1 & \cellcolor{realtime}79.9 & \cellcolor{realtime}55.8 & \cellcolor{realtime}34.2 & \cellcolor{realtime}18.4 & \cellcolor{realtime}82.0 & \cellcolor{realtime}66.9 & \cellcolor{realtime}52.4 & \cellcolor{realtime}37.5 & \cellcolor{realtime}23.9 & \cellcolor{realtime}70.6 & \cellcolor{realtime}59.8 & \cellcolor{realtime}48.7 & \cellcolor{realtime}38.0 & \cellcolor{realtime}27.0 \\

\rowcolor{tabbandb} & enc & \cellcolor{realtime}45.7 & \cellcolor{realtime}38.6 & \cellcolor{realtime}31.2 & \cellcolor{realtime}24.9 & \cellcolor{realtime}17.7 & 163.7 & 135.2 & 104.1 & \cellcolor{realtime}71.5 & \cellcolor{realtime}41.7 & 125.4 & 107.1 & \cellcolor{realtime}89.2 & \cellcolor{realtime}69.8 & \cellcolor{realtime}50.1 & 104.0 & \cellcolor{realtime}91.0 & \cellcolor{realtime}77.9 & \cellcolor{realtime}64.7 & \cellcolor{realtime}51.1 \\
\rowcolor{tabbandb}\multirow{-2}{*}{G-PCC (inter)} & dec & \cellcolor{realtime}36.4 & \cellcolor{realtime}30.6 & \cellcolor{realtime}24.3 & \cellcolor{realtime}18.9 & \cellcolor{realtime}14.3 & 129.4 & 103.7 & \cellcolor{realtime}77.6 & \cellcolor{realtime}51.5 & \cellcolor{realtime}29.5 & 101.9 & \cellcolor{realtime}85.8 & \cellcolor{realtime}69.7 & \cellcolor{realtime}54.3 & \cellcolor{realtime}38.8 & \cellcolor{realtime}85.2 & \cellcolor{realtime}73.6 & \cellcolor{realtime}62.4 & \cellcolor{realtime}50.4 & \cellcolor{realtime}39.0 \\

\rowcolor{tabbanda} & enc & 2249 & 1796 & 1362 & 968.4 & 641.7 & 7322 & 5352 & 3561 & 2044 & 1011 & 5411 & 4325 & 3280 & 2249 & 1314 & 4908 & 4087 & 3276 & 2426 & 1700 \\
\rowcolor{tabbanda}\multirow{-2}{*}{TopNet$^\dagger$} & dec & 725K & 594K & 445K & 313K & 203K & 1.59M & 1.20M & 787K & 453K & 222K & 1.22M & 983K & 732K & 495K & 292K & 1.08M & 890K & 702K & 550K & 366K \\

\rowcolor{tabbandb} & enc & 97.0 & 81.0 & 64.0 & 51.0 & \cellcolor{realtime}41.0 & 246.0 & 187.0 & 129.0 & \cellcolor{realtime}84.0 & \cellcolor{realtime}50.0 & 204.0 & 168.0 & 132.0 & \cellcolor{realtime}96.0 & \cellcolor{realtime}66.0 & 186.0 & 158.0 & 132.0 & 103.0 & \cellcolor{realtime}78.0 \\
\rowcolor{tabbandb}\multirow{-2}{*}{RENO} & dec & 88.0 & 76.0 & 62.0 & \cellcolor{realtime}50.0 & \cellcolor{realtime}37.0 & 242.0 & 186.0 & 128.0 & \cellcolor{realtime}81.0 & \cellcolor{realtime}46.0 & 205.0 & 170.0 & 134.0 & \cellcolor{realtime}97.0 & \cellcolor{realtime}65.0 & 178.0 & 151.0 & 124.0 & \cellcolor{realtime}99.0 & \cellcolor{realtime}73.0 \\

\rowcolor{tabbanda} & enc & 404.8 & 341.1 & 282.2 & 227.8 & 176.6 & 1025 & 773.6 & 537.1 & 339.8 & 208.5 & 869.0 & 697.5 & 559.0 & 403.9 & 278.6 & 784.4 & 671.4 & 555.7 & 441.9 & 325.1 \\
\rowcolor{tabbanda}\multirow{-2}{*}{AnyPcc} & dec & 387.7 & 318.6 & 263.1 & 203.9 & 156.2 & 1043 & 787.3 & 543.9 & 350.1 & 198.7 & 846.1 & 684.8 & 527.8 & 376.1 & 251.6 & 789.6 & 660.9 & 545.1 & 415.4 & 317.1 \\

\rowcolor{tabbandb} & enc & 77.0 & 64.0 & 51.0 & \cellcolor{realtime}42.0 & \cellcolor{realtime}33.0 & 204.0 & 155.0 & 110.0 & \cellcolor{realtime}73.0 & \cellcolor{realtime}48.0 & 164.0 & 135.0 & 106.0 & \cellcolor{realtime}79.0 & \cellcolor{realtime}54.0 & 152.0 & 129.0 & 107.0 & \cellcolor{realtime}85.0 & \cellcolor{realtime}64.0 \\
\rowcolor{tabbandb}\multirow{-2}{*}{ELiCv1} & dec & 73.0 & 62.0 & \cellcolor{realtime}49.0 & \cellcolor{realtime}41.0 & \cellcolor{realtime}33.0 & 207.0 & 157.0 & 111.0 & \cellcolor{realtime}69.0 & \cellcolor{realtime}42.0 & 170.0 & 139.0 & 109.0 & \cellcolor{realtime}82.0 & \cellcolor{realtime}54.0 & 150.0 & 126.0 & 107.0 & \cellcolor{realtime}82.0 & \cellcolor{realtime}62.0 \\
\midrule
\rowcolor{tabbanda} & enc & \cellcolor{realtime}31.2 & \cellcolor{realtime}26.2 & \cellcolor{realtime}22.6 & \cellcolor{realtime}18.6 & \cellcolor{realtime}15.2 & \cellcolor{realtime}95.3 & \cellcolor{realtime}78.1 & \cellcolor{realtime}59.9 & \cellcolor{realtime}41.6 & \cellcolor{realtime}25.5 & \cellcolor{realtime}69.9 & \cellcolor{realtime}60.7 & \cellcolor{realtime}51.0 & \cellcolor{realtime}40.8 & \cellcolor{realtime}29.3 & \cellcolor{realtime}61.0 & \cellcolor{realtime}54.7 & \cellcolor{realtime}48.0 & \cellcolor{realtime}41.2 & \cellcolor{realtime}33.2 \\
\rowcolor{tabbanda}\multirow{-2}{*}{ELiCv2 ($C{=}32$ intra)} & dec & \cellcolor{realtime}34.5 & \cellcolor{realtime}29.9 & \cellcolor{realtime}25.3 & \cellcolor{realtime}21.0 & \cellcolor{realtime}17.0 & 113.6 & \cellcolor{realtime}92.4 & \cellcolor{realtime}70.8 & \cellcolor{realtime}49.4 & \cellcolor{realtime}30.9 & \cellcolor{realtime}80.3 & \cellcolor{realtime}69.2 & \cellcolor{realtime}58.1 & \cellcolor{realtime}45.5 & \cellcolor{realtime}32.5 & \cellcolor{realtime}71.3 & \cellcolor{realtime}62.9 & \cellcolor{realtime}55.4 & \cellcolor{realtime}46.9 & \cellcolor{realtime}37.1 \\

\rowcolor{tabbandb} & enc & \cellcolor{realtime}34.8 & \cellcolor{realtime}31.2 & \cellcolor{realtime}27.6 & \cellcolor{realtime}23.5 & \cellcolor{realtime}19.6 & 101.1 & \cellcolor{realtime}85.0 & \cellcolor{realtime}67.5 & \cellcolor{realtime}47.5 & \cellcolor{realtime}31.0 & \cellcolor{realtime}75.2 & \cellcolor{realtime}66.7 & \cellcolor{realtime}57.8 & \cellcolor{realtime}46.9 & \cellcolor{realtime}34.2 & \cellcolor{realtime}65.6 & \cellcolor{realtime}59.5 & \cellcolor{realtime}53.5 & \cellcolor{realtime}47.4 & \cellcolor{realtime}38.4 \\
\rowcolor{tabbandb}\multirow{-2}{*}{ELiCv2 ($C{=}32$ inter)} & dec & \cellcolor{realtime}40.3 & \cellcolor{realtime}36.0 & \cellcolor{realtime}31.2 & \cellcolor{realtime}25.9 & \cellcolor{realtime}21.4 & 118.3 & \cellcolor{realtime}97.8 & \cellcolor{realtime}76.3 & \cellcolor{realtime}52.6 & \cellcolor{realtime}33.6 & \cellcolor{realtime}89.1 & \cellcolor{realtime}77.9 & \cellcolor{realtime}66.2 & \cellcolor{realtime}52.4 & \cellcolor{realtime}37.6 & \cellcolor{realtime}77.8 & \cellcolor{realtime}69.8 & \cellcolor{realtime}62.1 & \cellcolor{realtime}53.7 & \cellcolor{realtime}43.3 \\

\rowcolor{tabbanda} & enc & 55.6 & \cellcolor{realtime}48.4 & \cellcolor{realtime}42.2 & \cellcolor{realtime}34.6 & \cellcolor{realtime}28.3 & 199.7 & 174.2 & 138.2 & \cellcolor{realtime}94.6 & \cellcolor{realtime}55.1 & 138.8 & 126.1 & 111.5 & \cellcolor{realtime}90.9 & \cellcolor{realtime}64.3 & 116.4 & 107.8 & \cellcolor{realtime}98.8 & \cellcolor{realtime}87.7 & \cellcolor{realtime}71.8 \\
\rowcolor{tabbanda}\multirow{-2}{*}{ELiCv2 ($C{=}64$ intra)} & dec & 57.7 & 51.4 & \cellcolor{realtime}43.7 & \cellcolor{realtime}35.9 & \cellcolor{realtime}28.9 & 218.9 & 192.3 & 156.9 & 109.7 & \cellcolor{realtime}62.2 & 148.1 & 133.8 & 117.1 & \cellcolor{realtime}94.7 & \cellcolor{realtime}66.5 & 125.4 & 115.6 & 104.7 & \cellcolor{realtime}91.9 & \cellcolor{realtime}74.3 \\

\rowcolor{tabbandb} & enc & 59.8 & 54.6 & \cellcolor{realtime}48.1 & \cellcolor{realtime}40.2 & \cellcolor{realtime}33.2 & 208.6 & 184.3 & 149.6 & 102.1 & \cellcolor{realtime}60.4 & 147.1 & 135.3 & 121.6 & \cellcolor{realtime}99.0 & \cellcolor{realtime}70.2 & 122.7 & 114.9 & 106.3 & \cellcolor{realtime}95.9 & \cellcolor{realtime}78.3 \\
\rowcolor{tabbandb}\multirow{-2}{*}{ELiCv2 ($C{=}64$ inter)} & dec & 65.1 & 58.7 & 50.6 & \cellcolor{realtime}42.2 & \cellcolor{realtime}34.4 & 230.7 & 199.1 & 161.7 & 109.8 & \cellcolor{realtime}62.6 & 164.3 & 148.1 & 130.8 & 105.9 & \cellcolor{realtime}73.3 & 134.3 & 124.1 & 113.6 & 101.6 & \cellcolor{realtime}81.9 \\

\bottomrule
\end{tabular}%
}

\end{table}

Table~\ref{tab:latency-all} reports end-to-end latency on Jetson Thor.
The $C{=}32$ intra model is faster than RENO, ELiCv1, and AnyPcc across
all evaluated datasets and bit-depths.
On nuScenes, the $C{=}32$ inter model encodes and decodes a 16-bit frame
in $34.8$\,ms and $40.3$\,ms, respectively, meeting the $50$\,ms budget
for 20Hz operation. At 14 bits, these times decrease to $27.6$\,ms and
$31.2$\,ms. RENO, ELiCv1, and AnyPcc require $97.0$\,ms, $77.0$\,ms,
and $404.8$\,ms for 16-bit encoding, respectively. ELiCv2 is thus the
only evaluated learned codec to sustain 20Hz at the finest resolution.

At 16 bits, $C{=}32$ inter encoding takes $101.1$\,ms on SemanticKITTI,
$75.2$\,ms on Ford, and $65.6$\,ms on V2X-Seq. Temporal conditioning
adds only $3.6$--$5.8$\,ms over intra encoding across the four datasets.
Increasing width to $C{=}64$ improves compression at additional runtime
cost: nuScenes 16-bit inter encoding rises from $34.8$ to $59.8$\,ms.
We therefore use $C{=}32$ as the deployment-oriented configuration and
$C{=}64$ as the rate-oriented configuration. Power, energy, and memory
measurements appear in Appendices~\ref{app:power} and~\ref{app:memory}.

\subsection{Ablation Studies}
\label{sec:experiments-ablation}

\subsubsection{Discrete Entropy Interface}
\label{sec:experiments-ablation-entropy}

Table~\ref{tab:coder-stage-mean} compares continuous-CDF arithmetic
coding~\citep{witten1987arithmetic} and rANS against discrete-table indexed
rANS for the $C{=}32$ intra backbone. We
average the four datasets at bit-depth 16 after the shared network prediction.
Per-dataset results are in Appendix~\ref{app:entropy-details}.

\begin{table}[!ht]
\centering
\footnotesize
\setlength{\tabcolsep}{5pt}
\renewcommand{\arraystretch}{1.1}
\caption{
Entropy-coding ablation on Jetson Thor, averaged equally over the four
datasets. \emph{Prep.} denotes per-voxel CDF construction for continuous
coding and top-1 table selection for indexed rANS. \emph{D2H} denotes
device-to-host transfer from the GPU to the CPU.
}
\label{tab:coder-stage-mean}
\begin{tabular}{lrrrrrr}
\toprule
& \multicolumn{4}{c}{Encoding time (ms/frame)} & & \\
\cmidrule(lr){2-5}
Entropy coder & Prep. & D2H & Code & Total & $\times$ ours & bpp \\
\midrule
\rowcolor{tabbanda}
Arithmetic coding
& 38.24 & 2.33 & 14.27 & 54.84 & $5.0\times$ & 12.7872 ($-1.11\%$) \\
\rowcolor{tabbandb}
rANS
& 38.24 & 2.33 & 6.42 & 46.98 & $4.3\times$ & 12.7882 ($-1.10\%$) \\
\rowcolor{tabbanda}
Indexed rANS
& 5.75 & 0.80 & 4.05 & 10.61 & -- & 12.9271 \\
\bottomrule
\end{tabular}
\end{table}

Replacing arithmetic coding with rANS leaves the dominant $38.24$\,ms
CDF construction unchanged. Discrete table selection reduces preparation
to $5.75$\,ms and total entropy coding to $10.61$\,ms. D2H transfer also
falls because each voxel transfers one table index instead of a complete
CDF. Indexed rANS is $5.0\times$ faster than arithmetic coding and
$4.3\times$ faster than continuous-CDF rANS, at approximately $1.1\%$
rate cost. The gain comes primarily from the neural--coder interface.

\subsubsection{Kernel-Map Propagation}
\label{sec:experiments-ablation-kmap}

\begin{table}[!ht]
\centering
\small
\setlength{\tabcolsep}{5pt}
\renewcommand{\arraystretch}{1.05}
\caption{
Kernel-map construction latency on Jetson Thor for the
$3\!\times\!3\!\times\!3$ kernel used in ELiCv2.
Runtime includes all progressively constructed levels, with sparse
convolution excluded. Both methods produce identical kernel-maps.
}
\label{tab:kmap}
\begin{tabular}{lrrrr}
\toprule
Dataset & Voxels & Ours (ms) & Hashing (ms) & Speed-up \\
\midrule
\rowcolor{tabbanda}
SemanticKITTI & 120{,}826 & 1.69 & 5.78 & $3.42\times$ \\
\rowcolor{tabbandb}
Ford          &  80{,}002 & 1.46 & 5.16 & $3.54\times$ \\
\rowcolor{tabbanda}
nuScenes      &  29{,}824 & 0.70 & 2.80 & $4.00\times$ \\
\rowcolor{tabbandb}
V2X-Seq       &  63{,}416 & 1.27 & 4.51 & $3.55\times$ \\
\midrule
\rowcolor{tabbanda}
Mean & & 1.28 & 4.56
& $\mathbf{3.63\times}$ \\
\bottomrule
\end{tabular}
\end{table}

Table~\ref{tab:kmap} compares our progressive builder with TorchSparse++'s
hash-based method. Both produce identical $3\!\times\!3\!\times\!3$ kernel-maps,
but propagation reduces mean latency from $4.56$ to $1.28$\,ms.
The arithmetic mean of the per-dataset speedups is $3.63\times$. The $3.28$\,ms saving recovers $6.6\%$ of the $50$\,ms
budget for 20Hz operation, providing time for temporal conditioning.
This gain comes from eliminating coordinate search at finer levels.
Results for larger kernels appear in Appendix~\ref{app:kmap-runtime}.

%% file: 5_outro.tex
\section{Conclusion and Limitations}
\label{sec:conclusion}

ELiCv2 improves embedded LiDAR geometry compression by redesigning
the entropy-coding interface and exploiting the hierarchy exposed
by progressive occupancy coding. The entropy model directly selects
a coder-ready integer CDF, while the sparse engine derives finer-level
kernel-maps from decoded occupancy and parent kernel-maps.
Together, these mechanisms eliminate per-voxel CDF materialization
and finer-level coordinate search. ELiCv2 uses the recovered
computation budget for lightweight ego-motion-aware temporal
conditioning. Across four LiDAR datasets, the $C{=}32$ inter model
achieves a mean BD-rate reduction of $30.97\%$ over G-PCC.
On nuScenes, it sustains 20Hz encoding and decoding on an NVIDIA
Jetson Thor. These results demonstrate that reducing computation
at the entropy and sparse-processing interfaces can support
effective temporal prediction under real-time embedded constraints.

The current implementation performs neural inference in FP32,
leaving opportunities to reduce computation and memory through
integer quantization. Temporal conditioning relies on accurate
ego poses and sufficient local overlap with the previous decoded
frame. Its benefit may diminish with pose errors, dynamic objects,
and fine resolutions where LiDAR samples no longer align locally.
Future work will investigate deterministic low-precision inference
and temporal conditioning that remains lightweight while becoming
more robust to these cases.

%% file: app_split.tex
\section{Octant-Partition Ablation}
\label{app:split}

Each 8-bit occupancy symbol is coded as two sequential 4-bit partial symbols.
This operation partitions the eight bits of the occupancy symbol rather than
the voxel set itself. Each bit corresponds to one octant of the
$2\!\times\!2\!\times\!2$ child grid, so the partition determines which decoded
octant bits are available when predicting the second partial symbol. We
compare the octant-parity partition used by ELiCv2 with a planar partition
corresponding to the two nibbles of the occupancy byte. The planar partition
codes $\{4,5,6,7\}$ before $\{0,1,2,3\}$, which groups the bits associated
with the two planes along the $z$ axis. Our partition codes
$\{1,2,4,7\}$ before $\{0,3,5,6\}$. Every octant represented in the second
partial symbol is therefore face-adjacent to three octants represented in the
decoded first partial symbol instead of one.

\begin{figure}[h]
\centering
\includegraphics[width=0.8\textwidth]{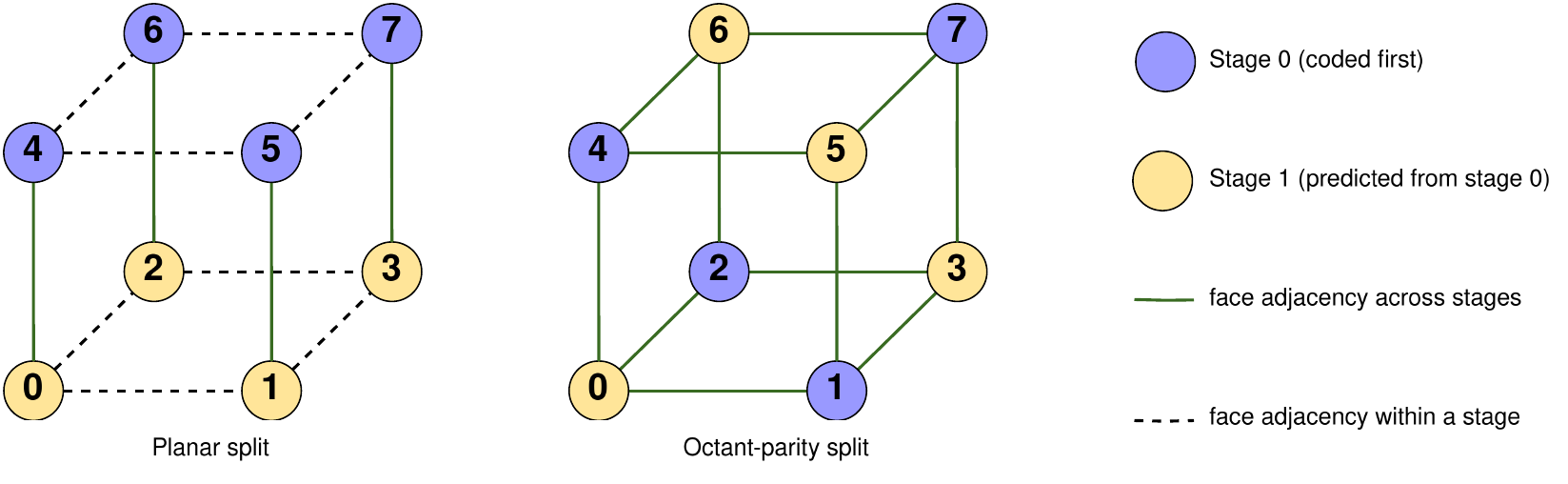}
\caption{Comparison of planar and octant-parity partitions of the eight bits
in an occupancy symbol. The diagram visualizes the octant associated with each
bit and does not indicate a partition of the voxel set. Stage 0 is coded first,
and Stage 1 is predicted from Stage 0. Solid edges connect
face-adjacent octants across stages, while dashed edges connect octants within
the same stage. Under octant parity, every Stage 1 octant has three
face-adjacent Stage 0 neighbors.}
\label{fig:octant-splits}
\end{figure}

Figure~\ref{fig:octant-splits} visualizes the two partitions and their
face-adjacency structure.

Table~\ref{tab:split-ablation} compares the two partitions with all other
settings fixed. Both variants use the same channel width, kernel size,
training schedule, and full-test evaluation procedure. Rate is computed with
the deployed top-1 table selection by summing the predicted code lengths and
normalizing by the number of occupied voxels at $\mathrm{posQ}=1$. This common
evaluation path isolates the occupancy partition from differences in the
bitstream implementations of the two code trees.

Octant parity achieves a lower rate at all 20 dataset and bit-depth
combinations. The reduction ranges from $0.062$ to $0.246$ bpp and averages
$0.125$ bpp. The mean reduction is largest on Ford at $0.189$ bpp, followed by
SemanticKITTI at $0.112$ bpp, V2X-Seq at $0.101$ bpp, and nuScenes at
$0.098$ bpp. These results support exposing spatially adjacent occupancy to
the second stage. The comparison covers two natural partitions at one channel
width and kernel size and does not claim optimality over all possible
partitions.

\begin{table}[t]
\centering
\footnotesize
\setlength{\tabcolsep}{3.8pt}
\renewcommand{\arraystretch}{1.08}
\caption{Rate comparison of octant-parity and planar occupancy splits for the
$C{=}32$ intra model with a $3\!\times\!3\!\times\!3$ kernel after $300$k training
steps. Values are bpp on the complete test split. Positive $\Delta$ indicates
that the planar split requires more bits.}
\label{tab:split-ablation}
\begin{tabular}{llrrrrr}
\toprule
Dataset & Split & 16-bit & 15-bit & 14-bit & 13-bit & 12-bit \\
\midrule
\rowcolor{tabbanda}
& Octant parity & \textbf{11.048} & \textbf{8.659} & \textbf{6.445} & \textbf{4.302} & \textbf{2.493} \\
\rowcolor{tabbanda}
& Planar & 11.144 & 8.783 & 6.584 & 4.424 & 2.571 \\
\rowcolor{tabbanda}
\multirow{-3}{*}{SemanticKITTI} & $\Delta$ & $+0.096$ & $+0.124$ & $+0.139$ & $+0.123$ & $+0.078$ \\
\midrule
\rowcolor{tabbandb}
& Octant parity & \textbf{12.912} & \textbf{10.377} & \textbf{8.075} & \textbf{6.021} & \textbf{4.095} \\
\rowcolor{tabbandb}
& Planar & 13.025 & 10.523 & 8.276 & 6.267 & 4.332 \\
\rowcolor{tabbandb}
\multirow{-3}{*}{Ford} & $\Delta$ & $+0.114$ & $+0.146$ & $+0.201$ & $+0.246$ & $+0.238$ \\
\midrule
\rowcolor{tabbanda}
& Octant parity & \textbf{10.878} & \textbf{8.528} & \textbf{6.561} & \textbf{4.883} & \textbf{3.481} \\
\rowcolor{tabbanda}
& Planar & 10.940 & 8.616 & 6.675 & 5.001 & 3.592 \\
\rowcolor{tabbanda}
\multirow{-3}{*}{nuScenes} & $\Delta$ & $+0.062$ & $+0.088$ & $+0.113$ & $+0.118$ & $+0.111$ \\
\midrule
\rowcolor{tabbandb}
& Octant parity & \textbf{15.950} & \textbf{13.171} & \textbf{10.570} & \textbf{8.244} & \textbf{6.246} \\
\rowcolor{tabbandb}
& Planar & 16.052 & 13.262 & 10.648 & 8.325 & 6.398 \\
\rowcolor{tabbandb}
\multirow{-3}{*}{V2X-Seq} & $\Delta$ & $+0.103$ & $+0.092$ & $+0.077$ & $+0.081$ & $+0.153$ \\
\bottomrule
\end{tabular}
\end{table}

%% file: app_entropy.tex
\section{Per-Dataset Entropy-Interface Ablation}
\label{app:entropy-details}

We report the per-dataset measurements underlying the averaged entropy-coding
ablation in Table~\ref{tab:coder-stage-mean}. All methods share the same network
prediction. The continuous baseline forms a probability-weighted mixture of
the codebook distributions, whereas indexed rANS uses the distribution
selected by hard top-1 row and column indices. The comparison therefore
isolates the interface between the shared network prediction and the entropy
coder.

A conventional entropy model produces a different continuous categorical
distribution for every voxel, whereas an integer entropy coder consumes
discrete cumulative frequencies. The interface must therefore convert each
voxel distribution into a complete monotonic integer CDF before coding. This
work scales with both the number of voxels and the number of symbols, and the
resulting CDF values must all cross the neural--coder boundary. ELiCv2 instead
predicts the index of a coder-ready distribution from a shared codebook. Each
voxel exposes only one compact index, and the corresponding integer CDF is
retrieved from a table constructed once rather than materialized again for
every voxel.

\begin{table}[h]
\centering
\footnotesize
\setlength{\tabcolsep}{3.5pt}
\renewcommand{\arraystretch}{1.08}
\caption{Per-dataset entropy-coding ablation on Jetson Thor using ten random
frames per dataset at bit-depth 16 with the $C{=}32$ intra model. \emph{Prep.}
denotes per-voxel CDF construction for continuous coding and top-1 table
selection for indexed rANS. Negative rate differences indicate lower bpp than
indexed rANS.}
\label{tab:coder-stage}
\begin{tabular}{llrrrrrrr}
\toprule
& & \multicolumn{4}{c}{Encoding time (ms/frame)} &
& \multicolumn{2}{c}{Rate} \\
\cmidrule(lr){3-6}\cmidrule(lr){8-9}
Dataset & Entropy coder & Prep. & D2H & Code & Total & $\times$ ours & bpp & $\Delta$ rate \\
\midrule
\rowcolor{tabbanda}
& Arithmetic coding & 15.22 & 1.85 & 5.37 & 22.44 & $4.3\times$ & 11.0319 & $-1.43\%$ \\
\rowcolor{tabbanda}
& rANS & 15.22 & 1.85 & 2.52 & 19.59 & $3.8\times$ & 11.0338 & $-1.42\%$ \\
\rowcolor{tabbanda}
\multirow{-3}{*}{nuScenes} & Indexed rANS & 2.75 & 0.74 & 1.69 & 5.18 & -- & 11.1924 & -- \\
\midrule
\rowcolor{tabbandb}
& Arithmetic coding & 53.96 & 2.62 & 19.89 & 76.46 & $5.4\times$ & 10.1504 & $-1.11\%$ \\
\rowcolor{tabbandb}
& rANS & 53.96 & 2.62 & 8.98 & 65.55 & $4.6\times$ & 10.1508 & $-1.11\%$ \\
\rowcolor{tabbandb}
\multirow{-3}{*}{SemanticKITTI} & Indexed rANS & 7.80 & 0.81 & 5.54 & 14.15 & -- & 10.2644 & -- \\
\midrule
\rowcolor{tabbanda}
& Arithmetic coding & 45.62 & 2.48 & 16.76 & 64.85 & $5.3\times$ & 12.9944 & $-0.98\%$ \\
\rowcolor{tabbanda}
& rANS & 45.62 & 2.48 & 7.62 & 55.71 & $4.5\times$ & 12.9950 & $-0.98\%$ \\
\rowcolor{tabbanda}
\multirow{-3}{*}{Ford} & Indexed rANS & 6.67 & 0.81 & 4.83 & 12.32 & -- & 13.1233 & -- \\
\midrule
\rowcolor{tabbandb}
& Arithmetic coding & 38.17 & 2.37 & 15.06 & 55.60 & $5.2\times$ & 16.9722 & $-0.91\%$ \\
\rowcolor{tabbandb}
& rANS & 38.17 & 2.37 & 6.54 & 47.08 & $4.4\times$ & 16.9732 & $-0.91\%$ \\
\rowcolor{tabbandb}
\multirow{-3}{*}{V2X-Seq} & Indexed rANS & 5.79 & 0.83 & 4.15 & 10.77 & -- & 17.1282 & -- \\
\bottomrule
\end{tabular}
\end{table}

Table~\ref{tab:coder-stage} shows the same pattern on every dataset. Replacing
arithmetic coding with rANS reduces coding time but leaves per-voxel CDF
construction and transfer unchanged. The discrete entropy interface instead
predicts compact table indices directly, reducing both preparation and
device-to-host transfer before indexed rANS coding. This interface change
provides a $3.8\times$ to $5.4\times$ total speed-up for a rate difference of
$0.91\%$ to $1.43\%$.

%% file: app_table_size.tex
\section{Effect of the Table Codebook Size}
\label{app:table-size}

The codebook size $T$ controls the rate--storage trade-off
of the discrete entropy interface in Section~\ref{sec:method-table}.
We examine this trade-off to motivate the default choice of $T{=}32$.

\paragraph{Setup.}
We evaluate $T \in \{16,32,64\}$ for geometry-only intra-frame coding
on SemanticKITTI with $C{=}32$. All models use the same training recipe
and four BoE coding networks, with two coding stages per network.
Only the codebook size and selection-head width vary.
We report model rates on the complete test split at five bit depths,
using deterministic $\arg\max$ selection and the same normalisation
as in the main experiments.

\begin{table}[!ht]
\centering
\footnotesize
\setlength{\tabcolsep}{3.8pt}
\renewcommand{\arraystretch}{1.08}
\caption{Effect of the table codebook size on SemanticKITTI for the
geometry-only intra model ($C{=}32$, $K{=}3$). Values are model bpp on the
complete test split. Best rates are in \textbf{bold}. Negative $\Delta$
indicates rate savings from doubling $T$; differences are computed before
rounding. CDF storage accounts for four BoE coding networks and two stages
per network.}
\label{tab:table-size}
\begin{tabular}{lrrrrrrrr}
\toprule
& \multicolumn{5}{c}{Geometry rate (bpp)}
& \multicolumn{3}{c}{CDF storage} \\
\cmidrule(lr){2-6}\cmidrule(lr){7-9}
$T$ & 16-bit & 15-bit & 14-bit & 13-bit & 12-bit
& $T^2$ & Rows & Size \\
\midrule
\rowcolor{tabbanda}
$16$ & 11.258 & 8.845 & 6.599 & 4.416 & 2.560
& 256 & 2{,}048 & 68\,KiB \\
\rowcolor{tabbandb}
$32$ (default) & 11.048 & 8.659 & 6.445 & 4.302 & 2.493
& 1{,}024 & 8{,}192 & 272\,KiB \\
\rowcolor{tabbanda}
$64$ & \textbf{10.950} & \textbf{8.574} & \textbf{6.373}
& \textbf{4.252} & \textbf{2.463}
& 4{,}096 & 32{,}768 & 1.0625\,MiB \\
\midrule
\rowcolor{tabbandb}
$\Delta\,16\!\to\!32$
& $-0.210$ & $-0.186$ & $-0.154$ & $-0.114$ & $-0.067$ & & & \\
\rowcolor{tabbanda}
$\Delta\,32\!\to\!64$
& $-0.098$ & $-0.086$ & $-0.072$ & $-0.050$ & $-0.030$ & & & \\
\bottomrule
\end{tabular}
\end{table}

\paragraph{Results.}
Table~\ref{tab:table-size} shows diminishing rate gains as $T$ increases.
Doubling $T$ from $16$ to $32$ saves $0.067$--$0.210$\,bpp across the five
bit depths, whereas doubling it again saves only $0.030$--$0.098$\,bpp.
Each doubling quadruples CDF storage, since the coder stores
$4 \times 2 \times T^2$ distributions: $68$\,KiB, $272$\,KiB, and
$1.0625$\,MiB for the three configurations.
At $b{=}16$, reducing $T$ from $32$ to $16$ costs $0.210$\,bpp,
while increasing it to $64$ saves only $0.098$\,bpp.
We therefore use $T{=}32$ as a practical rate--storage compromise.
This ablation uses one training seed per configuration on a single dataset;
decoding latency at $T \neq 32$ was not measured.

%% file: app_kmap.tex
\section{Kernel-Map Reuse Derivation and Example}
\label{app:rulebook}

This appendix expands the kernel-map reuse of
Section~\ref{sec:method-kmap}. We derive how child-level connectivity can be
obtained directly from the parent kernel-map and decoded occupancy, give the
two lookup tables used by our implementation, and provide a worked example.
We denote the active parent voxels by $\mathcal{P}=\{\mathbf{p}_i\}$, their kernel-map
by $\mathbf{K}^{p}\in\mathbb{Z}^{N_p\times27}$, the child occupancy mask of
parent $i$ by the 8-bit integer $m_i\in\{0,\ldots,255\}$, and the resulting
child kernel-map by
$\mathbf{K}^{c}\in\mathbb{Z}^{N_c\times27}$.
We use $m_i[q]\in\{0,1\}$ to denote the bit associated with octant $q$.

\paragraph{Deterministic child indexing.}
Each parent voxel generates up to eight children. We denote an octant by the
zero-based scalar index $q=x+2y+4z$ for $(x,y,z)\in\{0,1\}^{3}$ and denote
its binary offset by $\boldsymbol{\beta}(q)=(x,y,z)$. Only children satisfying
$m_i[q]=1$ are instantiated. We assign active children contiguous indices
using
\begin{equation}
\operatorname{idx}(i,q)
=
\operatorname{base}(i)
+
\operatorname{popcount}\!\left(m_i \wedge (2^{q}-1)\right),
\qquad
\operatorname{base}(i)
=
\sum_{j<i}\operatorname{popcount}(m_j) ,
\label{eq:app-child-idx}
\end{equation}
where $\operatorname{base}(i)$ is an exclusive prefix sum over the numbers of active
children. Thus, the global index of every child is determined directly from
its parent index and occupancy mask, without constructing child coordinates.

\paragraph{Hierarchical connectivity.}
Consider a child of parent $\mathbf{p}$ at octant $q$ with coordinate
$\mathbf{x}=2\mathbf{p}+\boldsymbol{\beta}(q)$. For a
submanifold-convolution offset $\mathbf{d}\in\{-1,0,1\}^{3}$, its target
neighbor is $\mathbf{x}'=2\mathbf{p}+\boldsymbol{\beta}(q)+\mathbf{d}$.
We decompose this target as $\mathbf{x}'=2\mathbf{p}'+\boldsymbol{\gamma}$, where
\begin{equation}
\mathbf{t}
=
\left\lfloor\frac{\boldsymbol{\beta}(q)+\mathbf{d}}{2}\right\rfloor,
\qquad
\mathbf{p}'=\mathbf{p}+\mathbf{t},
\qquad
\boldsymbol{\gamma}=\boldsymbol{\beta}(q)+\mathbf{d}-2\mathbf{t}.
\label{eq:app-decomp}
\end{equation}
Here, $\mathbf{t}\in\{-1,0,1\}^{3}$ identifies the neighboring parent containing the
target, while $\boldsymbol{\gamma}\in\{0,1\}^{3}$ identifies the target child offset within
that parent. Therefore, a child-level neighbor can always be resolved by one
parent-level lookup followed by one occupancy test.

For the $3\!\times\!3\!\times\!3$ kernel, we index an offset
$\mathbf{d}_k=(d_x,d_y,d_z)$ by
\begin{equation}
k=(d_x+1)+3(d_y+1)+9(d_z+1),
\qquad k\in\{0,\ldots,26\}.
\label{eq:app-koffset}
\end{equation}
Since Equation~\eqref{eq:app-decomp} depends only on the source child octant $q$
and kernel index $k$, its result can be precomputed into two small lookup
tables,
\begin{equation}
T_{\mathrm{col}}(q,k)=k',
\qquad
T_{\mathrm{bit}}(q,k)=\operatorname{ind}(\boldsymbol{\gamma}),
\label{eq:app-luts}
\end{equation}
where $T_{\mathrm{col}}$ specifies which parent-level kernel-map column to
consult and $T_{\mathrm{bit}}$ specifies which child octant to test within
that neighboring parent. Here,
$\operatorname{ind}(x,y,z)=x+2y+4z$ converts a binary offset into its scalar
octant index. Tables~\ref{tab:app-tcol}
and~\ref{tab:app-tbit} provide the complete lookup tables.

\begin{table*}[t]
\centering
\caption{
Lookup table $T_{\mathrm{col}}(q,k)$ for the $3\!\times\!3\!\times\!3$
submanifold kernel. Each entry specifies the parent-level kernel-map column
containing the target child. Together with $T_{\mathrm{bit}}$ in
Table~\ref{tab:app-tbit}, it completely determines child-level connectivity.
}
\label{tab:app-tcol}
\resizebox{\textwidth}{!}{
\begin{tabular}{c|ccccccccccccccccccccccccccc}
\toprule
$q \backslash k$
& 0 & 1 & 2 & 3 & 4 & 5 & 6 & 7 & 8
& 9 & 10 & 11 & 12 & 13 & 14 & 15 & 16 & 17
& 18 & 19 & 20 & 21 & 22 & 23 & 24 & 25 & 26 \\
\midrule
0 & 0 & 1 & 1 & 3 & 4 & 4 & 3 & 4 & 4 & 9 & 10 & 10 & 12 & 13 & 13 & 12 & 13 & 13 & 9 & 10 & 10 & 12 & 13 & 13 & 12 & 13 & 13 \\
1 & 1 & 1 & 2 & 4 & 4 & 5 & 4 & 4 & 5 & 10 & 10 & 11 & 13 & 13 & 14 & 13 & 13 & 14 & 10 & 10 & 11 & 13 & 13 & 14 & 13 & 13 & 14 \\
2 & 3 & 4 & 4 & 3 & 4 & 4 & 6 & 7 & 7 & 12 & 13 & 13 & 12 & 13 & 13 & 15 & 16 & 16 & 12 & 13 & 13 & 12 & 13 & 13 & 15 & 16 & 16 \\
3 & 4 & 4 & 5 & 4 & 4 & 5 & 7 & 7 & 8 & 13 & 13 & 14 & 13 & 13 & 14 & 16 & 16 & 17 & 13 & 13 & 14 & 13 & 13 & 14 & 16 & 16 & 17 \\
4 & 9 & 10 & 10 & 12 & 13 & 13 & 12 & 13 & 13 & 9 & 10 & 10 & 12 & 13 & 13 & 12 & 13 & 13 & 18 & 19 & 19 & 21 & 22 & 22 & 21 & 22 & 22 \\
5 & 10 & 10 & 11 & 13 & 13 & 14 & 13 & 13 & 14 & 10 & 10 & 11 & 13 & 13 & 14 & 13 & 13 & 14 & 19 & 19 & 20 & 22 & 22 & 23 & 22 & 22 & 23 \\
6 & 12 & 13 & 13 & 12 & 13 & 13 & 15 & 16 & 16 & 12 & 13 & 13 & 12 & 13 & 13 & 15 & 16 & 16 & 21 & 22 & 22 & 21 & 22 & 22 & 24 & 25 & 25 \\
7 & 13 & 13 & 14 & 13 & 13 & 14 & 16 & 16 & 17 & 13 & 13 & 14 & 13 & 13 & 14 & 16 & 16 & 17 & 22 & 22 & 23 & 22 & 22 & 23 & 25 & 25 & 26 \\
\bottomrule
\end{tabular}
}
\end{table*}

\begin{table*}[t]
\centering
\caption{
Lookup table $T_{\mathrm{bit}}(q,k)$ for the $3\!\times\!3\!\times\!3$
submanifold kernel. Each entry specifies the target child octant within the
neighboring parent selected by $T_{\mathrm{col}}$.
}
\label{tab:app-tbit}
\resizebox{\textwidth}{!}{
\begin{tabular}{c|ccccccccccccccccccccccccccc}
\toprule
$q \backslash k$
& 0 & 1 & 2 & 3 & 4 & 5 & 6 & 7 & 8
& 9 & 10 & 11 & 12 & 13 & 14 & 15 & 16 & 17
& 18 & 19 & 20 & 21 & 22 & 23 & 24 & 25 & 26 \\
\midrule
0 & 7 & 6 & 7 & 5 & 4 & 5 & 7 & 6 & 7 & 3 & 2 & 3 & 1 & 0 & 1 & 3 & 2 & 3 & 7 & 6 & 7 & 5 & 4 & 5 & 7 & 6 & 7 \\
1 & 6 & 7 & 6 & 4 & 5 & 4 & 6 & 7 & 6 & 2 & 3 & 2 & 0 & 1 & 0 & 2 & 3 & 2 & 6 & 7 & 6 & 4 & 5 & 4 & 6 & 7 & 6 \\
2 & 5 & 4 & 5 & 7 & 6 & 7 & 5 & 4 & 5 & 1 & 0 & 1 & 3 & 2 & 3 & 1 & 0 & 1 & 5 & 4 & 5 & 7 & 6 & 7 & 5 & 4 & 5 \\
3 & 4 & 5 & 4 & 6 & 7 & 6 & 4 & 5 & 4 & 0 & 1 & 0 & 2 & 3 & 2 & 0 & 1 & 0 & 4 & 5 & 4 & 6 & 7 & 6 & 4 & 5 & 4 \\
4 & 3 & 2 & 3 & 1 & 0 & 1 & 3 & 2 & 3 & 7 & 6 & 7 & 5 & 4 & 5 & 7 & 6 & 7 & 3 & 2 & 3 & 1 & 0 & 1 & 3 & 2 & 3 \\
5 & 2 & 3 & 2 & 0 & 1 & 0 & 2 & 3 & 2 & 6 & 7 & 6 & 4 & 5 & 4 & 6 & 7 & 6 & 2 & 3 & 2 & 0 & 1 & 0 & 2 & 3 & 2 \\
6 & 1 & 0 & 1 & 3 & 2 & 3 & 1 & 0 & 1 & 5 & 4 & 5 & 7 & 6 & 7 & 5 & 4 & 5 & 1 & 0 & 1 & 3 & 2 & 3 & 1 & 0 & 1 \\
7 & 0 & 1 & 0 & 2 & 3 & 2 & 0 & 1 & 0 & 4 & 5 & 4 & 6 & 7 & 6 & 4 & 5 & 4 & 0 & 1 & 0 & 2 & 3 & 2 & 0 & 1 & 0 \\
\bottomrule
\end{tabular}
}
\end{table*}

\FloatBarrier

\paragraph{Kernel-map propagation.}
Given the two lookup tables, the child kernel-map is generated directly from
the parent kernel-map and occupancy masks. For an active child
$u=\operatorname{idx}(i,q)$ and kernel index $k$, we first obtain
\begin{equation}
k'=T_{\mathrm{col}}(q,k),
\qquad
q'=T_{\mathrm{bit}}(q,k),
\qquad
j=\mathbf{K}^{p}[i,k'].
\end{equation}
The corresponding child-level entry is then
\begin{equation}
\mathbf{K}^{c}[u,k]
=
\begin{cases}
\operatorname{idx}(j,q'),
&
\text{if } j\geq0 \text{ and } m_j[q']=1,\\
-1,
&
\text{otherwise.}
\end{cases}
\label{eq:app-kmap-fill}
\end{equation}
Thus, each child-level connection requires only fixed table lookups, a
parent-kernel-map access, an occupancy test, and child indexing. No child
coordinate hashing or dynamic neighbor search is required.

\paragraph{Worked example.}
Consider the three parent voxels
\begin{equation}
\mathbf{p}_0=(0,0,0),\qquad
\mathbf{p}_1=(1,0,0),\qquad
\mathbf{p}_2=(1,0,1),
\end{equation}
with occupancy masks
\begin{equation}
m_0=0b00100010,\qquad
m_1=0b00000011,\qquad
m_2=0b00010001.
\end{equation}
The active octants are therefore $\{1,5\}$, $\{0,1\}$, and $\{0,4\}$,
respectively. Equation~\eqref{eq:app-child-idx} assigns the resulting children
the contiguous indices shown in Table~\ref{tab:app-child-index-example}.

\begin{table}[t]
\centering
\small
\caption{
Child indices for the worked example. The global child ordering is obtained
directly from the parent occupancy masks using Equation~\eqref{eq:app-child-idx}.
}
\label{tab:app-child-index-example}
\begin{tabular}{ccccc}
\toprule
Parent $i$ & Active $q$ & $\operatorname{base}(i)$ &
$\operatorname{idx}(i,q)$ & Child coordinate \\
\midrule
0 & 1 & 0 & 0 & $(1,0,0)$ \\
0 & 5 & 0 & 1 & $(1,0,1)$ \\
1 & 0 & 2 & 2 & $(2,0,0)$ \\
1 & 1 & 2 & 3 & $(3,0,0)$ \\
2 & 0 & 4 & 4 & $(2,0,2)$ \\
2 & 4 & 4 & 5 & $(2,0,3)$ \\
\bottomrule
\end{tabular}
\end{table}

\FloatBarrier

As one concrete propagation, consider $\mathbf{K}^{c}[2,21]$.
From Table~\ref{tab:app-child-index-example}, child $u=2$ corresponds to
$(i,q)=(1,0)$. Reading row $q=0$ and column $k=21$ of
Tables~\ref{tab:app-tcol} and~\ref{tab:app-tbit} gives
\begin{equation}
T_{\mathrm{col}}(0,21)=12,
\qquad
T_{\mathrm{bit}}(0,21)=5.
\end{equation}
The parent kernel-map gives
\begin{equation}
j=\mathbf{K}^{p}[1,12]=0,
\end{equation}
so the target lies in parent $\mathbf{p}_0$. Since $m_0[5]=1$, child octant $5$
exists, and Equation~\eqref{eq:app-child-idx} gives
\begin{equation}
\operatorname{idx}(0,5)=1.
\end{equation}
Therefore,
\begin{equation}
\mathbf{K}^{c}[2,21]=1.
\end{equation}
The same procedure is applied independently to every active child and kernel
offset to construct the complete child kernel-map.

\paragraph{Equivalence to coordinate-based construction.}
Kernel-map reuse changes only how connectivity is obtained, not the
connectivity itself. For any child coordinate
$\mathbf{x}=2\mathbf{p}+\boldsymbol{\beta}(q)$ and kernel offset $\mathbf{d}$,
Equation~\eqref{eq:app-decomp} uniquely identifies the parent containing
$\mathbf{x}+\mathbf{d}$ and the corresponding child octant. Equation~\eqref{eq:app-kmap-fill}
therefore produces the same child-level neighbor relation as explicit
coordinate-based search, provided that the parent kernel-map is identical.
The proposed method replaces repeated coordinate discovery with deterministic
connectivity propagation while leaving the sparse convolution unchanged.

%% file: app_large_kmap.tex
\section{Kernel-Map Construction with Larger Kernels}
\label{app:kmap-runtime}

Our kernel-map propagation in Section~\ref{sec:method-kmap} exploits a property
specific to progressive coarse-to-fine occupancy coding. At each level, the
active voxels are not an arbitrary coordinate set. They are deterministically
generated by expanding the previous level according to the decoded child
occupancy. Consequently, once the parent kernel-map and child occupancy are
known, the connectivity of the finer level can be propagated without
reconstructing it through coordinate-based neighbor search.

While this advantage arises from the progressive coding structure, it is not
restricted to the $3\!\times\!3\!\times\!3$ kernel used in ELiCv2. The same
parent-to-child connectivity propagation can be constructed for larger
neighborhoods by extending the corresponding offset lookup tables. We verify
this by evaluating $3\!\times\!3\!\times\!3$, $5\!\times\!5\!\times\!5$,
$7\!\times\!7\!\times\!7$, and $9\!\times\!9\!\times\!9$ kernels.

\paragraph{Measurement.}
We isolate kernel-map construction from sparse convolution and compare our
progressive builder with the hash-based builder of TorchSparse++. For each
frame, runtime is accumulated over all progressively constructed octree
levels from bit-depth $3$ to the level preceding the leaf. Bit-depth $2$ is
excluded because it has no parent kernel-map and is constructed identically
by both methods. All experiments are conducted on Jetson Thor at bit-depth
16 ($\mathrm{posQ}=4$). Both implementations were verified to produce identical
kernel-maps at every evaluated level and kernel size.

Table~\ref{tab:kmap-all} reports construction latency for all evaluated kernel
sizes and datasets.

\begin{table}[t]
\centering
\small
\setlength{\tabcolsep}{4pt}
\renewcommand{\arraystretch}{1.05}
\caption{
Kernel-map construction latency on Jetson Thor for different kernel sizes.
Our method propagates connectivity across progressively decoded octree levels,
whereas the baseline reconstructs each kernel-map using TorchSparse++'s
hash-based builder. Both methods produce identical kernel-maps.
}
\label{tab:kmap-all}
\begin{tabular}{llrrrr}
\toprule
Kernel & Dataset & Voxels & Ours (ms) & Hashing (ms) & Speed-up \\
\midrule

\rowcolor{tabbanda}& SemanticKITTI & 120{,}826 & 1.69 & 5.78 & $3.42\times$ \\
\rowcolor{tabbanda}& Ford          &  80{,}002 & 1.46 & 5.16 & $3.54\times$ \\
\rowcolor{tabbanda}& nuScenes      &  29{,}824 & 0.70 & 2.80 & $4.00\times$ \\
\rowcolor{tabbanda}& V2X-Seq       &  63{,}416 & 1.27 & 4.51 & $3.55\times$ \\
\rowcolor{tabbanda}\multirow{-5}{*}{$3\!\times\!3\!\times\!3$}& Mean & & 1.28 & 4.56
& $\mathbf{3.63\times}$ \\
\midrule

\rowcolor{tabbandb}& SemanticKITTI & 120{,}826 & 7.58 & 21.17 & $2.79\times$ \\
\rowcolor{tabbandb}& Ford          &  80{,}002 & 6.15 & 17.62 & $2.86\times$ \\
\rowcolor{tabbandb}& nuScenes      &  29{,}824 & 2.24 & 6.59 & $2.94\times$ \\
\rowcolor{tabbandb}& V2X-Seq       &  63{,}416 & 4.97 & 14.16 & $2.85\times$ \\
\rowcolor{tabbandb}\multirow{-5}{*}{$5\!\times\!5\!\times\!5$}& Mean & & 5.24 & 14.88
& $\mathbf{2.86\times}$ \\
\midrule

\rowcolor{tabbanda}& SemanticKITTI & 120{,}826 & 25.84 & 54.02 & $2.09\times$ \\
\rowcolor{tabbanda}& Ford          &  80{,}002 & 20.59 & 45.06 & $2.19\times$ \\
\rowcolor{tabbanda}& nuScenes      &  29{,}824 &  6.88 & 15.55 & $2.26\times$ \\
\rowcolor{tabbanda}& V2X-Seq       &  63{,}416 & 16.82 & 36.60 & $2.18\times$ \\
\rowcolor{tabbanda}\multirow{-5}{*}{$7\!\times\!7\!\times\!7$}& Mean & & 17.53 & 37.81
& $\mathbf{2.18\times}$ \\
\midrule

\rowcolor{tabbandb}& SemanticKITTI & 120{,}826 & 58.16 & 111.30 & $1.91\times$ \\
\rowcolor{tabbandb}& Ford          &  80{,}002 & 46.28 &  92.67 & $2.00\times$ \\
\rowcolor{tabbandb}& nuScenes      &  29{,}824 & 15.71 &  31.79 & $2.02\times$ \\
\rowcolor{tabbandb}& V2X-Seq       &  63{,}416 & 37.61 &  75.10 & $2.00\times$ \\
\rowcolor{tabbandb}\multirow{-5}{*}{$9\!\times\!9\!\times\!9$}& Mean & & 39.44 & 77.71
& $\mathbf{1.98\times}$ \\
\bottomrule
\end{tabular}
\end{table}

\FloatBarrier

\paragraph{Generalization across kernel sizes.}
Table~\ref{tab:kmap-all} shows that the computational advantage enabled by
progressive occupancy coding is not specific to the $3\!\times\!3\!\times\!3$
stencil. For the $3\!\times\!3\!\times\!3$ kernel used in ELiCv2, our builder
reduces mean construction latency from $4.56$ to $1.28$\,ms.
The arithmetic mean of the per-dataset speedups is $3.63\times$.
The same construction remains effective for larger neighborhoods, with
mean per-dataset speedups of $2.86\times$, $2.18\times$, and $1.98\times$
for $5\!\times\!5\!\times\!5$, $7\!\times\!7\!\times\!7$, and $9\!\times\!9\!\times\!9$ kernels,
respectively.

The relative gain decreases with kernel size because our builder explicitly
propagates the connectivity for all $K^3$ kernel offsets, whereas the
hash-based implementation amortizes part of its coordinate-map construction
across offsets. Nevertheless, the proposed method remains consistently faster
for every tested kernel size and dataset while producing exactly the same
kernel-maps.

These results clarify both the scope and the source of the proposed
optimization. Kernel-map propagation is enabled by the deterministic
parent--child structure inherent in progressive coarse-to-fine occupancy
coding, rather than by a particular convolution stencil. Therefore, within
such progressive codecs, the same principle can be applied beyond the
$3\!\times\!3\!\times\!3$ kernel used in ELiCv2 without reverting to coordinate-based
connectivity search.

%% file: app_inter_align.tex
\section{Ego-Motion Alignment and Reference Resolution}
\label{app:inter-align}

\begin{figure}[t]
    \centering
    \includegraphics[width=\textwidth]{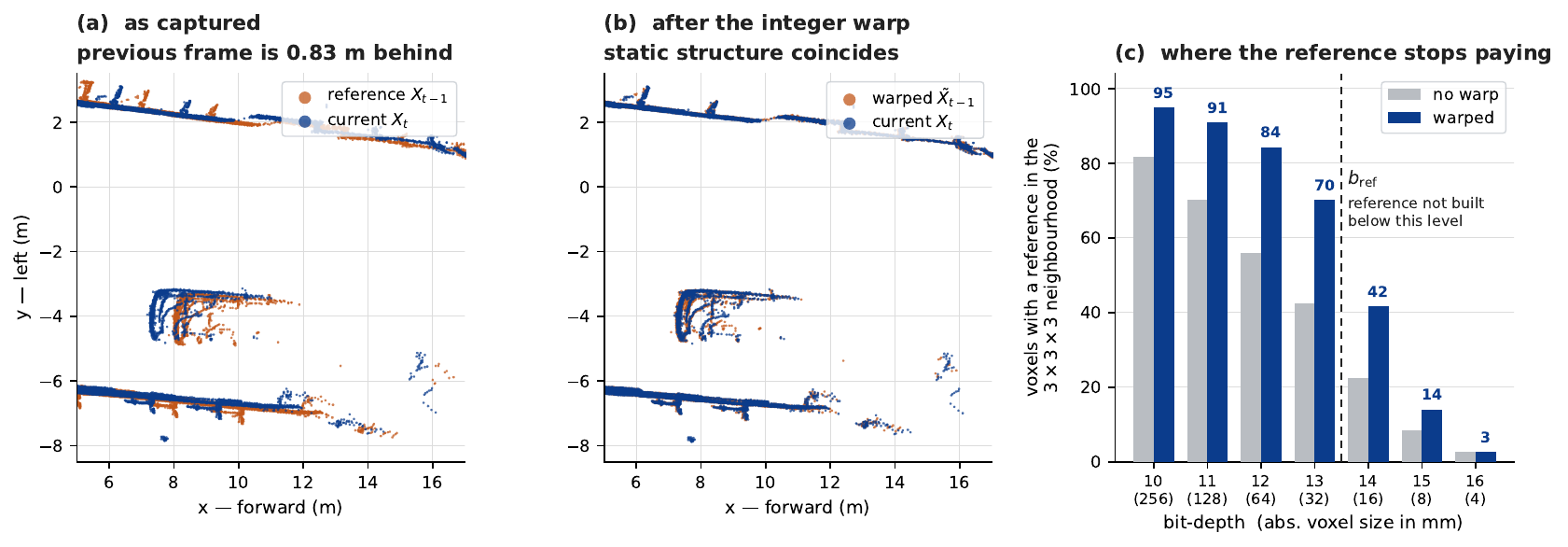}
    \caption{
    Effect of ego-motion alignment on the inter reference.
    (a) Previous and current frames before alignment.
    (b) The previous frame after the deterministic ego-motion warp.
    (c) Fraction of current voxels having an occupied reference within a
    $3\!\times\!3\!\times\!3$ neighborhood at each bit-depth.
    The dashed line denotes the reference cutoff $\ell_{\mathrm{ref}}=13$.
    }
    \label{fig:inter-align}
\end{figure}

Figure~\ref{fig:inter-align} visualizes the ego-motion alignment used by the
inter model and examines how its effectiveness changes with spatial resolution.
The example is taken from consecutive frames of SemanticKITTI sequence 18,
where the sensor translates by $0.826$\,m with a yaw change of
$0.82^\circ$.

\paragraph{Why alignment is necessary.}
Even moderate vehicle motion is large relative to the voxel resolution used
for inter prediction. In Figure~\ref{fig:inter-align}(a), the $0.826$\,m
translation corresponds to approximately $26$ voxels at bit-depth $13$
($32$\,mm cells), substantially displacing static structures between
consecutive frames. After applying the relative ego pose, these structures
become locally aligned, as shown in Figure~\ref{fig:inter-align}(b).
Ego-motion compensation therefore provides a useful temporal reference without
learned motion estimation or correspondence search.

\paragraph{Deterministic encoder--decoder alignment.}
To ensure identical reference construction at the encoder and decoder, the
relative rigid transform is quantized once into fixed-point parameters
$(\mathbf{R},\mathbf{t})$ and transmitted with each P-frame. Reference
coordinates are then generated entirely with integer arithmetic. Both sides
therefore construct bit-identical warped coordinates without independently
evaluating floating-point pose transformations.

\paragraph{Why the reference stops at $\ell_{\mathrm{ref}}=13$.}
The reference pyramid is used only for $\ell\leq \ell_{\mathrm{ref}}=13$,
corresponding to a $32$\,mm cell. This cutoff is defined by absolute bit-depth,
so its physical resolution is unchanged across operating points. Two effects
make finer references progressively less useful. First, the
$3\!\times\!3\!\times\!3$ neighborhood has a radius of one voxel, so its physical
coverage halves at each finer level. Second, consecutive LiDAR sweeps sample
different surface points, causing voxel-level correspondence itself to vanish
at sufficiently fine resolutions even after correct ego-motion alignment.

This behavior is visible in Figure~\ref{fig:inter-align}(c). For the shown
frame pair, alignment increases the local reference hit rate from $81.6\%$ to
$94.9\%$ at a $256$\,mm cell and from $42.4\%$ to $70.1\%$ at the
$32$\,mm cells of $\ell=13$. The aligned hit rate then drops to $41.7\%$ at
$16$\,mm and $14.0\%$ at $8$\,mm. More generally, on SemanticKITTI the
reference hit rate decreases from $63\%$ at $\ell=13$ to $32\%$ at $\ell=14$ and
$8.5\%$ at $\ell=15$. Increasing the neighborhood radius to two voxels recovers
much of the $\ell=14$ correspondence but provides little recovery at $\ell=15$,
indicating that the fine-scale degradation is not merely caused by a limited
search radius. We therefore restrict temporal conditioning to $\ell\leq13$,
where the warped reference retains reliable local correspondence.

\FloatBarrier
\paragraph{Sensitivity to ego-motion.}

\begin{figure}[t]
\centering
\includegraphics[width=\textwidth]{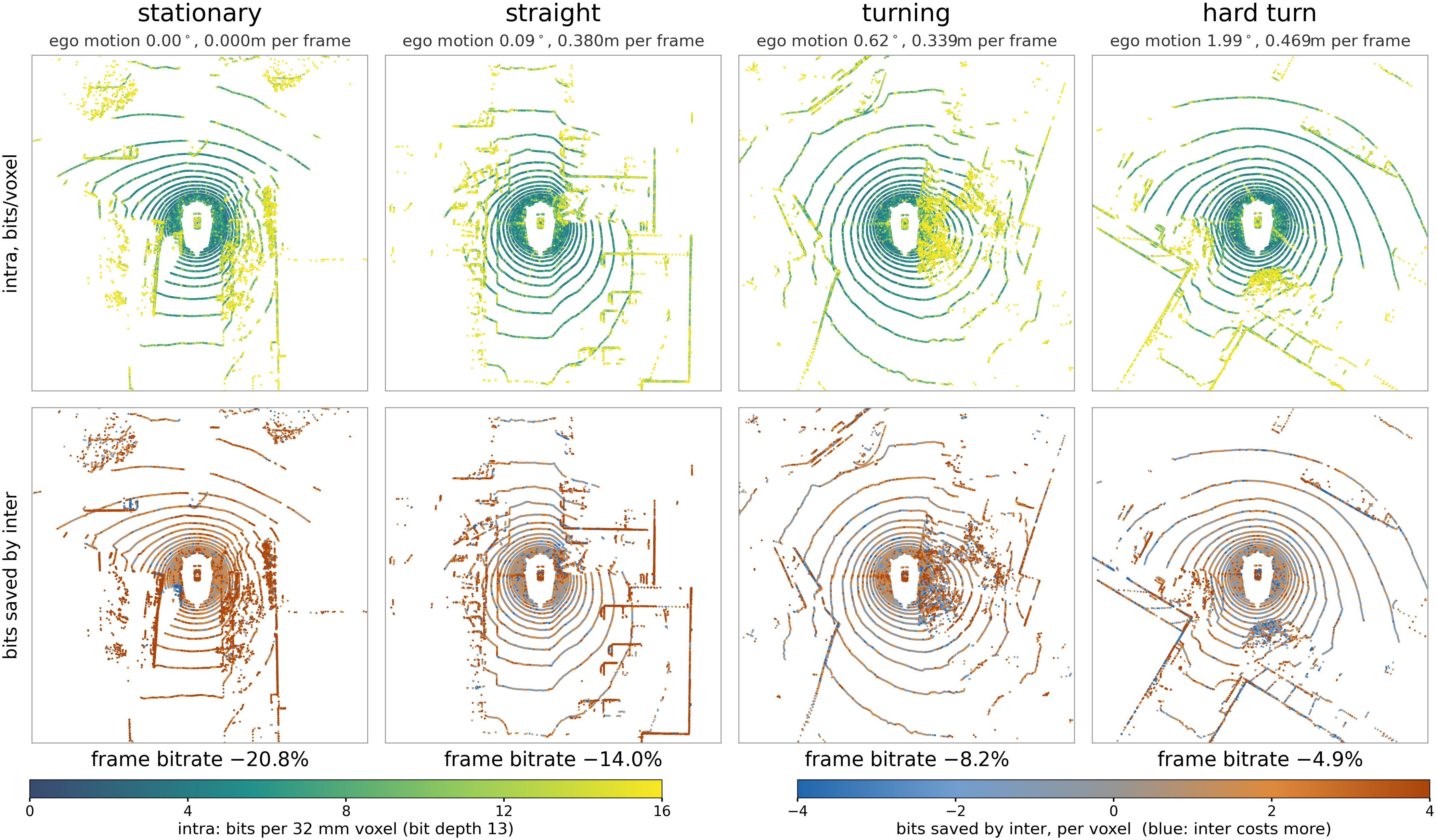}
\caption{Per-voxel intra coding cost (top) and inter-frame bit saving (bottom)
under four ego-motion regimes in nuScenes. Positive values in the bottom row
indicate savings from the aligned reference. Examples are from the same drive.}
\label{fig:motion-cases}
\end{figure}

Figure~\ref{fig:motion-cases} compares per-voxel coding behavior under four
representative ego-motion regimes.

\FloatBarrier

Figure~\ref{fig:motion-cases} illustrates variation in coding gains
across ego-motion regimes within the same drive, using the same
sensor and model. A nearly stationary frame obtains a $20.8\%$ rate
reduction, and straight driving with $0.380$\,m translation and $0.09^\circ$
rotation obtains $14.0\%$. Turning with similar translation ($0.339$\,m) but
$0.62^\circ$ rotation yields $8.2\%$, while a $1.99^\circ$ hard turn yields
$4.9\%$. Although ego-motion alignment compensates for rigid displacement,
changes in visibility and surface sampling can still reduce
correspondence between consecutive sweeps. These examples illustrate
motion-associated variation rather than isolate the effect of rotation
alone.

Despite this variation, $99.9\%$ of the $47{,}734$ nuScenes P-frames improve
over intra coding, with a mean saving of $13.1\%$. The reference gate can
suppress uninformative temporal context, avoiding a motion-dependent fallback.
These within-drive results characterize motion sensitivity. Cross-dataset gains
also depend on viewpoint consistency, scene dynamics, and intra-model strength.

%% file: app_why.tex
\section{Rationale for Ego-Motion Alignment}
\label{app:why-alignment}

Our inter model first aligns the previous LiDAR sweep using the sensor
ego-motion and then reads a local $3\!\times\!3\!\times\!3$ neighborhood.
An alternative, adopted by existing learned inter point cloud codecs, is to
leave the two frames unaligned and let a learned operator search for
cross-frame correspondence. We explain why ELiCv2 instead uses explicit
ego-motion alignment for embedded LiDAR compression.

\paragraph{Cost of cross-frame correspondence.}

Without alignment, the displacement between consecutive vehicle-mounted LiDAR
frames can be much larger than a local sparse-convolution stencil.
As analyzed in Appendix~\ref{app:inter-align}, a $3\!\times\!3\!\times\!3$ window
would need to grow substantially to cover the displacement caused by typical
ego-motion. Existing inter-frame methods therefore employ wider or
data-dependent cross-frame operators.

Table~\ref{tab:crossframe-ops} summarizes representative designs.
D-DPCC~\citep{ddpcc} combines cross-frame features and performs
nearest-neighbor interpolation. \citet{akhtar2024} use a
$3\!\times\!3\!\times\!3$ geometry-aware sparse convolution, while
Unicorn~\citep{unicorn} uses a $9\!\times\!9\!\times\!9$ target convolution for its
LiDAR warping model. In contrast, ELiCv2 removes the dominant rigid
displacement before temporal conditioning, allowing the reference operator
to remain local.

\begin{table}[t]
\centering
\small
\setlength{\tabcolsep}{4pt}
\renewcommand{\arraystretch}{1.1}
\caption{Cross-frame operators in learned inter point cloud codecs.
\emph{Taps} denotes the number of reference positions accessed per output
position for fixed-stencil operators.}
\label{tab:crossframe-ops}
\begin{tabular}{llrl}
\toprule
Method & Cross-frame operator & Taps & Application \\
\midrule
\rowcolor{tabbanda}
D-DPCC~\citep{ddpcc}
    & Conv.\ + $k$-NN interpolation & $27/3^\dagger$ & Single scale \\
\rowcolor{tabbandb}
Akhtar et al.~\citep{akhtar2024}
    & GSConv $3\!\times\!3\!\times\!3$ & $27$ & Once \\
\rowcolor{tabbanda}
Unicorn~\citep{unicorn}
    & Target conv.\ $9\!\times\!9\!\times\!9$ & $729$ & Multi-scale \\
\midrule
\rowcolor{tabbandb}
Ours
    & Ego warp + local gather & $27$ & $\ell\leq13$ \\
\bottomrule
\end{tabular}

\vspace{2pt}
\footnotesize
$^\dagger$ The interpolation uses a $k{=}3$ nearest-neighbor query rather than
three fixed stencil accesses.
\end{table}

\FloatBarrier

Sparse convolution requires a kernel-map that specifies which voxels are
neighbors. Within one frame, ELiCv2 can propagate this map through the known
octree hierarchy. A cross-frame operator instead connects voxels from two
different frames, whose coordinates change for every frame pair. It must
therefore construct a new cross-frame kernel-map each time. A wide cross-frame
operator is costly for two reasons. It accesses more neighbors and must also
build the connectivity between the two frames.

\FloatBarrier
\paragraph{Illustrative cost of wide kernel-map construction.}

To illustrate the cost of wide sparse neighborhoods, we refer to the
within-frame kernel-map construction benchmark in
Appendix~\ref{app:kmap-runtime}. For a $9\!\times\!9\!\times\!9$ kernel,
construction takes $39.44$\,ms with our progressive builder and
$77.71$\,ms with coordinate hashing, averaged over the four datasets
and accumulated across the evaluated octree levels.

These measurements demonstrate that wide kernel-map construction can
incur substantial overhead. They do not estimate the cost of the
cross-frame operators in Table~\ref{tab:crossframe-ops}, whose active
supports, scales, and map-construction procedures differ from this
benchmark.

ELiCv2 instead uses ego-motion alignment followed by a local
$3\!\times\!3\!\times\!3$ occupancy gather. This temporal path requires no
cross-frame sparse-convolution kernel-map, although reference warping
and local lookup still incur computation.

\paragraph{Benefits of ego-motion alignment.}

For vehicle-mounted LiDAR, much of the displacement between consecutive sweeps
is caused by the rigid motion of the sensor itself. Instead of asking the
network to rediscover this displacement, we transmit the quantized relative
ego transform and warp the reference onto the current coordinate system.
The transform requires only $96$ bytes per P-frame and is applied using the
integer warp in Equation~\eqref{eq:warp}.

After alignment, temporal conditioning requires only a local
$3\!\times\!3\!\times\!3$ occupancy lookup. This gives three practical advantages.
First, the dominant rigid displacement is removed before inference, avoiding
the need for a wide correspondence search. Second, the temporal path does not
require an additional cross-frame sparse-convolution backbone or its associated
kernel-map construction. Third, the fixed-point transform guarantees that the
encoder and decoder construct the same aligned reference exactly.

This choice is specific to driving LiDAR. Ego-motion cannot compensate for independently
moving objects, whereas learned flow or correspondence models can represent
such motion. For deformable point cloud sequences, this flexibility can be
essential. In vehicle-mounted LiDAR, however, ego-motion is readily available
and removes a large deterministic component of inter-frame displacement at
negligible side-information cost. Appendix~\ref{app:inter-align} visualizes
the resulting alignment and correspondence rates and evaluates the inter gain
across different ego-motion regimes.

Our design therefore uses the network primarily to model residual temporal
dependency rather than recover a rigid transform already known to the platform.
This design allows ELiCv2 to obtain inter-frame gains using a local reference
gate without motion search or complex temporal processing.

%% file: app_energy.tex
\section{Energy per Frame}
\label{app:power}

\paragraph{Measurement protocol.}
We measure the marginal energy per frame of learned codecs on Jetson Thor
using the module-level INA238 sensor sampled at 200\,Hz. Encoding and decoding
are measured separately over 30-frame runs. For each run, we subtract the idle
power measured immediately beforehand and integrate the remaining power over
the I/O-excluded execution time, then divide by the number of processed frames.
This metric measures the additional energy consumed by coding on an
already-powered platform; it excludes the platform's idle energy.
Table~\ref{tab:power-learned} reports the arithmetic mean across the four
datasets at bit-depth 16.

\begin{table}[h]
\centering
\small
\setlength{\tabcolsep}{10pt}
\renewcommand{\arraystretch}{1.1}
\caption{Marginal energy per frame at bit-depth 16 ($\mathrm{posQ}=4$) on
Jetson Thor, averaged equally over four datasets. Idle energy is excluded.
Lower values are better; the lowest value in each column is in \textbf{bold}.}
\label{tab:power-learned}
\begin{tabular}{lrr}
\toprule
Codec & Encode (mJ/frame) & Decode (mJ/frame) \\
\midrule
\rowcolor{tabbanda}
RENO                       & 3{,}496 & 3{,}607 \\
\rowcolor{tabbandb}
ELiCv1                     & 3{,}423 & 3{,}631 \\
\rowcolor{tabbanda}
AnyPcc                     & 21{,}345 & 21{,}246 \\
\midrule
\rowcolor{tabbandb}
ELiCv2 $C{=}32$ intra         & \textbf{1{,}695} & \textbf{2{,}118} \\
\rowcolor{tabbanda}
ELiCv2 $C{=}32$ inter         & 1{,}970 & 2{,}375 \\
\rowcolor{tabbandb}
ELiCv2 $C{=}64$ intra         & 4{,}626 & 5{,}782 \\
\rowcolor{tabbanda}
ELiCv2 $C{=}64$ inter         & 4{,}997 & 6{,}110 \\
\bottomrule
\end{tabular}
\end{table}

\FloatBarrier

\paragraph{Energy efficiency.}
The $C{=}32$ intra model has the lowest encoding and decoding energy among
the codecs in Table~\ref{tab:power-learned}, requiring $1{,}695$ and
$2{,}118$\,mJ/frame, respectively. With temporal conditioning, the $C{=}32$
inter model requires $1{,}970$\,mJ/frame for encoding and
$2{,}375$\,mJ/frame for decoding. Its encoding energy is approximately
$44\%$ lower than RENO and $42\%$ lower than ELiCv1, while its decoding
energy is approximately $34\%$ and $35\%$ lower, respectively.
Both $C{=}32$ variants also require substantially less energy than AnyPcc,
which consumes more than $21{,}000$\,mJ/frame for either operation.

\paragraph{Cost of temporal conditioning and model capacity.}
For $C{=}32$, switching from intra to inter coding increases encoding energy
by $275$\,mJ/frame ($16.2\%$) and decoding energy by $257$\,mJ/frame
($12.1\%$). This additional energy accompanies the improvement in mean
BD-rate reduction from $26.66\%$ to $30.97\%$ reported in
Table~\ref{tab:bdrate-all}. Thus, temporal conditioning improves compression
while retaining lower energy per frame than RENO and ELiCv1.

Increasing the width to $C{=}64$ further improves compression but raises
energy consumption. The $C{=}64$ inter model requires $4{,}997$\,mJ/frame
for encoding and $6{,}110$\,mJ/frame for decoding, approximately $2.5\times$
and $2.6\times$ the corresponding $C{=}32$ inter values. These measurements
support using $C{=}32$ when energy efficiency is a priority and $C{=}64$
when the additional compression gain justifies the higher energy cost.

%% file: app_memory.tex
\section{Memory Footprint}
\label{app:memory}

We measure GPU memory on Jetson Thor after two warm-up passes. Activation
memory is the increase from the resident baseline to the maximum allocated
memory during coding. Table~\ref{tab:memory} reports encoding and decoding on
SemanticKITTI at bit-depth 16, the most memory-intensive operating point in
our evaluation.

\begin{table}[t]
\centering
\small
\setlength{\tabcolsep}{10pt}
\renewcommand{\arraystretch}{1.1}
\caption{Activation memory for SemanticKITTI encoding and decoding at
bit-depth 16 on Jetson Thor. Memory is measured after warm-up. Lower values
are better.}
\label{tab:memory}
\begin{tabular}{lrr}
\toprule
Method & Encode (MB) & Decode (MB) \\
\midrule
\rowcolor{tabbanda}
RENO & 181.0 & 209.5 \\
\rowcolor{tabbandb}
ELiCv1 & \textbf{138.0} & 165.1 \\
\rowcolor{tabbanda}
AnyPcc & 912.0 & 863.2 \\
\midrule
\rowcolor{tabbandb}
ELiCv2 ($C{=}32$, intra) & 174.5 & \textbf{118.8} \\
\rowcolor{tabbanda}
ELiCv2 ($C{=}32$, inter) & 229.0 & 209.0 \\
\rowcolor{tabbandb}
ELiCv2 ($C{=}64$, intra) & 239.0 & 181.7 \\
\rowcolor{tabbanda}
ELiCv2 ($C{=}64$, inter) & 325.0 & 301.0 \\
\bottomrule
\end{tabular}
\end{table}

\FloatBarrier

The $C{=}32$ intra decoder uses the least memory among the evaluated learned
codecs, reducing activation memory by $28\%$ relative to ELiCv1, $43\%$
relative to RENO, and $86\%$ relative to AnyPcc. ELiCv2 also uses less memory
for decoding than encoding across all four configurations. Inter coding retains
the temporal reference and therefore increases memory, while the $C{=}64$
configurations trade additional memory for compression efficiency. All codecs listed in Table~\ref{tab:memory} fit on Jetson Thor, so this comparison reflects headroom for
concurrent perception and planning workloads and portability to smaller
devices rather than device compatibility.

\paragraph{Joint deployment trade-off.}
Memory should be considered together with compression efficiency and latency.
At the same SemanticKITTI bit-depth 16 operating point, the $C{=}32$ inter
model uses $229.0$\,MB for encoding and requires $101.1$\,ms, compared with
$138.0$\,MB and $204.0$\,ms for ELiCv1 and $181.0$\,MB and $246.0$\,ms for
RENO. The modest increase in encoding memory therefore approximately halves
latency while improving the mean BD-rate reduction from $27.54\%$ for ELiCv1
and $17.02\%$ for RENO to $30.97\%$. Its decoding memory is also comparable to
RENO at $209.0$ versus $209.5$\,MB. The $C{=}64$ inter model offers a
rate-oriented option with a stronger $35.22\%$ mean BD-rate reduction at a
higher memory cost. Together, these results show that ELiCv2 provides a
favorable compression--latency--memory trade-off rather than optimizing one
deployment metric in isolation.